\documentclass{emulateapj} 
\usepackage{epsfig}
\usepackage{subfigure}
\usepackage{amssymb}
\usepackage{color}

\newcommand{\bn}{\begin{enumerate}}
\newcommand{\en}{\end{enumerate}}
\newcommand{\bi}{\begin{itemize}}
\newcommand{\ei}{\end{itemize}}

\newcommand{\Msun}{\rm M_{\odot}}
\newcommand{\Mach}{\mathcal M}
\newcommand{\K}{\rm K}

\def\gtorder{\mathrel{\raise.3ex\hbox{$>$}\mkern-14mu
    \lower0.6ex\hbox{$\sim$}}}
\def\ltorder{\mathrel{\raise.3ex\hbox{$<$}\mkern-14mu
    \lower0.6ex\hbox{$\sim$}}}

\begin{document}
\shorttitle{Supermassive Black Hole Formation at High $z$ via Direct Collapse}
\shortauthors{Choi, Shlosman and Begelman}

\title{Supermassive Black Hole Formation at High Redshifts via Direct Collapse:\\
               Physical Processes in the Early Stage}

\author{Jun-Hwan Choi\altaffilmark{1}, Isaac Shlosman\altaffilmark{1}, \& Mitchell C. 
Begelman\altaffilmark{2}$^,$\altaffilmark{3}}
\altaffiltext{1}{Department of Physics \& Astronomy, University of Kentucky, Lexington, 
KY 40506-0055, USA, {\tt jhchoi@pa.uky.edu; shlosman@pa.uky.edu}}
\altaffiltext{2}{JILA, University of Colorado and National Institute of Standards and Technology, 
    440 UCB, Boulder, CO 80309-0440, USA}
\altaffiltext{3}{Department of Astrophysical and Planetary Sciences, 391UCB, University of 
    Colorado, Boulder, CO 80309-0391, USA, {\tt mitch@jila.colorado.edu}}

\begin{abstract}
We use numerical simulations to explore whether direct collapse can lead to the formation of supermassive 
black hole (SMBH) seeds at high redshifts. Using the adaptive mesh refinement code ENZO, we follow the 
evolution of gas within slowly tumbling dark matter (DM) halos of $M_{\rm vir}\sim 2\times 10^8\,M_\odot$ 
and $R_{\rm vir}\sim 1\,$kpc. For our idealized simulations, we adopt cosmologically motivated DM and 
baryon density profiles and angular momentum distributions.  
Our principal goal is to understand how the collapsing flow overcomes the centrifugal barrier and whether it 
is subject to fragmentation which can potentially lead to star formation, decreasing the seed SMBH mass.  
We find that the collapse proceeds from inside out and leads either to a central runaway or to off-center 
fragmentation.  A disk-like configuration is formed inside the centrifugal barrier, growing via accretion.  
For models with a more cuspy DM distribution, the gas collapses more and experiences a bar-like 
perturbation and a {\it central runaway} on scales of $\ltorder 1-10$\,pc.  We have followed this inflow 
down to $\sim 10^{-4}$\,pc ($\sim 10$\,AU), where it is estimated to become optically thick.
The flow remains isothermal and the specific angular momentum, $j$, is efficiently transferred by 
gravitational torques in a cascade of nested bars. This cascade is triggered by finite perturbations from 
the large-scale mass distribution and by gas self-gravity, and 
supports a self-similar, disk-like collapse where the axial ratios remain constant.  The mass accretion rate 
shows a global minimum on scales of $\sim 1-10$\,pc at the time of the central runaway. In the collapsing 
phase, virial supersonic turbulence develops and  fragmentation is damped. Models with progressively larger 
initial DM cores evolve similarly, but the timescales become longer. In models with more organized initial 
rotation --- when the rotation of spherical shells is constrained to be coplanar --- a torus forms on 
scales $\sim 20-50$\,pc outside the disk, and appears to be supported by turbulent motions driven by 
accretion from the outside. The overall evolution of the models depends on the competition between two 
timescales, corresponding to the onset of the central runaway and of off-center fragmentation. In models 
with less organized rotation --- when the rotation of spherical shells is randomized (but the total angular 
momentum remains unchanged) --- the torus is greatly weakened, the central accretion timescale is shortened, 
and off-center fragmentation is suppressed ---  triggering the central runaway even in previously `stable' 
models.  The resulting seed SMBH masses is found in the range 
$M_\bullet\sim 2\times 10^4\,M_\odot-2\times 10^6\,M_\odot$, substantially higher than the mass range of 
Population\,III remnants. We argue that the above upper limit on $M_\bullet$ appears to be more realistic,
and lies close to the cutoff mass of detected SMBHs. Corollaries of this 
model include a possible correlation between SMBH and DM halo masses, and similarity between the SMBH and 
halo mass functions, at time of formation.  
\end{abstract}


\section{Introduction}
\label{sec:intro}

Observational evidence points to most large galaxies hosting supermassive black holes (SMBHs) in their 
centers \citep[e.g.,][]{Ferrares:05} and to the possible coevolution of galaxies and SMBHs, whether causal 
\citep[e.g.,][]{Gebhardtetal:00,Ferrares.Merritt:00,Tremaine.etal:02} or not \citep{Jahnke.Maccio:11}. 
A growing number of QSOs have been found at $z > 6$, including a bright QSO at $z \gtorder 7$ 
\citep[e.g.,][]{Fan.etal:03,Mortlock.etal:11}. Hence, some 
SMBHs with $M_\bullet\gtorder\, $few$\times 
10^9\,\Msun$ had already formed before the universe was $\sim 700$\,Myr old, raising the issue of how such 
large black holes (BHs) could have grown so quickly.
  
The early formation and growth of SMBHs must be understood in a cosmological context.  Early SMBHs could 
have formed from remnants of the first generation of stars, Population III, which subsequently grew by gas
accretion and mergers 
\citep[e.g.,][]{Haiman.Loeb:01,Yoo.Miralda-Escude:04,Volonteri.Rees:06,Li.etal:07,Pelupessy.etal:07,Tanaka.Haiman:09}.  
Early studies of the initial
mass function (IMF) of these objects argued in favor of very massive $M_{\rm star} \sim 100-1,000\,\Msun$ 
objects \citep[e.g.,][]{Abel.etal:02,Bromm.etal:02,OShea.Norman:07}, but more recent work indicates a 
rather normal IMF \citep[e.g.,][]{Turk.etal:09,Clark.etal:11,Hosokawa.etal:11,Stacy.etal:12,Wise.etal:12}. 
Even if there were a sufficient supply of $100\,\Msun$ remnant BHs which grew via gas accretion at the 
Eddington rate with 10\% radiation efficiency, it would have taken them at least $\sim 7\times 10^8$\,yr 
to reach $\sim 10^9\,\Msun$ \citep{Salpeter:64}.  This timescale barely matches the available time required 
to make bright QSOs at $z\sim 6-7$.

Such efficient growth by accretion is unlikely, for two reasons. First, Population III stars in the 
vicinity of the seed, and the BH progenitor itself, ionize the surrounding matter, causing expansion of 
the H\,II region and likely suppression of subsequent accretion onto the SMBH 
\citep[e.g.,][]{Safranek.etal:12}.  Second,  the possible supernova explosion of a Pop\,III star can evacuate 
the ambient gas in the vicinity of the seed. 
This suggests that mergers must play an essential role in growing SMBHs from Population III seeds, but 
this is difficult as well. According to the cold dark matter (CDM) cosmology, 
dark matter (DM) halos frequently merge in hierarchical structure formation.  However, too frequent halo 
mergers would result in the formation of small  $N$-body systems of BH seeds, which would be prone to SMBH 
slingshot ejection(s) from the DM halo.  Given the relatively short time available for growing SMBH seeds to 
quasar masses, the long list of delaying processes can seriously hurt the Pop\,III seed scenario.

An appealing alternative is that early SMBHs formed via direct collapse of the halo gas at $z\sim 10-20$ 
\citep[e.g.,][]{Oh.Haman:02,Bromm.Loeb:03,Volonteri.Rees:05,Begelman.etal:06, 
Wise.etal:08,Levine.etal:08,Regan.Haehnelt:09,Begelman.Shlosman:09,Mayer.etal:10,Schleicher.etal:10,
Hosokawa.etal:11,Johnson.etal:11,Prieto.etal:13}. The direct collapse 
paradigm assumes that an SMBH seed, $M_\bullet\gtorder 10^5\,\Msun$, forms from the cold halo gas. This 
process favors a high density, high inflow rate environment in which star formation is suppressed.
A massive central object --- a supermassive star (SMS), forms at the center of the collapsing region. This 
object 
is powered by a combination of core nuclear burning and Kelvin-Helmholtz contraction 
\citep[e.g.,][]{Begelman.etal:06,Begelman.etal:08,Begelman:10}. After the stellar core collapses and forms 
the SMBH seed, its convective envelope is powered by the accretion onto this seed --- this configuration is 
termed a quasistar. Consequently, the seed BH, $\sim 100\,M_\odot$ at the beginning, can grow
to $\sim 10^5\,M_\odot$ in less than a few Myrs.

In this context, constraints on the accretion rate onto the SMBH become weaker, and additional problems 
anticipated in 
the Pop\,III seed scenario do not play a major role. In order to make this scenario work, the gas needs to 
inflow rapidly from $\sim$\,kpc to $\sim$\,100\,AU scales. The most probable site for such runaway gas 
collapse is expected in DM halos with virial temperatures $\gtorder 10^4$\,K. If the halo gas can cool only 
via atomic cooling, the necessary condition for the collapse to be triggered is for the gas temperature to 
be lower than the virial temperature of the host DM halo.

However, two caveats to the direct collapse scenario must be addressed \citep[e.g.,][]{Begelman.Shlosman:09}. 
First, the angular momentum barrier, in principle, can terminate the collapse well before it reaches $\sim 
\,100$\,AU scales. Owing to angular momentum conservation, collapse is halted when the rotational velocity 
reaches the circular velocity. In order to overcome this barrier, there should be a physical process to 
redistribute the gas angular momentum outward.  Second, gas could fragment during the collapse, depleting
the accretion stream by forming gas clumps and ultimately stars. The gas clumps could also 
disturb the accretion pattern. 

In this paper, we study the physical processes that accompany the initial and intermediate stages of gas 
collapse in a DM halo. We focus on the dynamical processes: spontaneous and induced symmetry breaking, 
nested gaseous bar formation, and the role of  supersonic turbulence.  Section~\ref{sec:theory} discusses 
the relevant processes.  
In Section~\ref{sec:method} we describe the numerical setups for a set of simulations intended to quantify 
these processes, and in Section~\ref{sec:simresult} we present our results.  We summarize and discuss our 
results in Section~\ref{sec:summary}.

\section{Theory}
\label{sec:theory}

\subsection{Angular Momentum Transfer}
\label{sec:AngMom}

It is generally understood that local viscous transport mechanisms are inefficient on galactic scales. Even 
on scales of a few parsecs, their characteristic timescales are prohibitively long and angular momentum 
transport therefore requires nonlocal mechanisms, e.g., large-scale magnetic fields \citep[e.g.,][]{Blandford.Payne:82}, or by gravitational torques. Turbulent motions, e.g., driven by the gravitational
collapse \citep[e.g.,][]{Hoyle:53,Klessen.Hennebelle:10,Sur.etal:10,Vazquez-Semadeni:10} and other
mechanisms, can amplify the seed magnetic fields \citep[e.g.,][for a review]{Balbus.Hawley:98,Subramanian:10}.
On larger 
spatial scales, the angular momentum redistribution most likely depends on long-range gravitational torques 
\citep[e.g.,][]{Lynden-Bel.Kalnajs:72,Shlosman.etal:89,Shlosman.etal:90}. When angular momentum is conserved, 
gas collapse is quickly stopped by the centrifugal barrier. Prior to collapse within DM halos, the gas can 
exhibit the same angular momentum distribution as the DM.
The typical spin acquired by a halo during maximal expansion is given by a dimensionless 
parameter $\lambda = J/\sqrt{2}M_{\rm vir}v_{\rm c}R_{\rm vir}$, where $J$ is the halo angular momentum, 
and $M_{\rm vir}$ and $v_{\rm c}$ are its virial mass and the circular velocity at the virial radius 
$R_{\rm vir}$ \citep[e.g.,][]{Peebles:69,Barnes.Efstathiou:87,Bullock.etal:01}. The mean spin of DM halos 
is $\lambda\sim 0.05$. If the gravitational potential of the gas dominates, this $\lambda$ will bring the 
gas to a centrifugal barrier when it has collapsed by a factor of $\sim 100$.  When the DM potential 
dominates, the collapse will stop after a decade in radius \citep[e.g., ][and refs. therein]{Mo:98,Shlosman:13}.

Inflow beyond the centrifugal barrier requires substantial and continuous angular momentum loss by the gas. 
What are the options?  If the gas disk becomes nonaxisymmetric, gravitational torques between the inner 
and outer gas, as well as between the gas and DM, can drain angular momentum away from the collapsing gas. 
The lowest nonaxisymmetric modes, $m = 1-3$, prevail in both the gas and the DM, because their rise times 
are shortest. The $m=1$ mode requires perturbation of the center
of mass of the gaseous disk --- this is not always possible. 
The next fastest growing mode is $m=2$ --- the bar-like mode. Typically it will 
have the highest amplitude, in the absence of $m=1$. For the inflow to extend to smaller spatial scales, 
the gas disk must be self-gravitating. Under these conditions, a cascade of bars can be 
maintained over a substantial dynamic range in radius. Contraction of the gas via 
such an avalanche can be related to self-organized criticality \citep{Lich.Lieb:95}. An analogy can be 
made between the chaos driven when the bar exceeds a certain strength and a sandpile whose slope is 
increased until the sand slides off \citep{Shlosman:05}.  

The bar cascade can be triggered in a number of ways.  First, if the disk becomes self-gravitating and 
cold, its axial symmetry is known to be broken spontaneously when the ratio of bulk kinetic energy to 
absolute potential energy, $T/|W|$, is larger than a certain critical value. The gas bar exerts torque on 
the gas disk and transfers angular momentum outward 
\citep[e.g.,][]{Shlosman.etal:89,Shlosman.etal:90,Englmaier.Shlosman:04}. This process allows the 
collapsing gas to cross 
the centrifugal barrier and to continue the collapse to smaller scales, if angular momentum is continuously 
extracted. The threshold value for gaseous bar formation depends also on the shape of the gas distribution 
and is $\alpha\equiv 0.5f T/|W| \geq 0.34$ \citep{Christodoulou.etal:95}, where $f=1$ for disks and 
$f=\frac{2}{3}$ for spheres.  This instability is spontaneous and does not require any trigger --- it
will grow exponentially from any infinitesimal perturbation.

Several additional effects can drive a cascade through finite amplitude perturbations. First, the shape 
and dynamics 
of a self-gravitating gas disk will respond to the potential of the DM halo in which it is embedded.  DM 
halos are known to be triaxial 
\citep[e.g.,][]{Allgood.etal:06,Berentzen.etal:06,Berentzen.Shlosman:06,Heller.etal:07},
and therefore exert gravitational torques on the disk, which will respond with low Fourier modes. Second,
finite bar perturbations can also be driven by tidal effects of massive DM and baryon clumps, i.e.,
halo substructure
\citep[e.g.,][]{Romano.etal:08} and from mergers. In addition, asymmetry in the background gravitational 
potential can be generated by the spontaneous break of axial symmetry in the collapsing flow, DM, gas or 
both. It is well-known that nonradial perturbations can grow in supersonic accretion flows onto compact 
objects \citep[e.g.,][]{Hunter:62,Garlick:79,Moncrief:80,Goldreich.etal:96,Lai.Goldreich:00}, without any 
assistance from self-gravity. Gas collapse inside an axisymmetric DM halo is similar --- in 
Section~\ref{sec:collapse} of the present work, we show the 
appearance of nonradial perturbations in the equatorial plane of the disk, although we start from idealized 
conditions of an isolated halo and embedded baryons which possess axial symmetry.   
In a comprehensive cosmological simulation, all these factors would provide {\it finite} amplitude 
perturbations that do not require the $\alpha$-parameter to cross the threshold.

At the spatial scale on which the gas disk forms, the potential contribution from the gas is already 
comparable with that of the DM,  and the gas cooling time is shorter than the dynamical time.  
These conditions imply that 
the collapsing gas becomes self-gravitating and nearly isothermal, and the density profile evolves toward a 
singular isothermal sphere\footnote{Throughout this paper, we use $R$ for spherical radii, and $r$ for 
cylindrical radii.}  
$\rho\propto R^{-2}$.  Before the gas reaches the centrifugal barrier, we can assume it flows in with the 
free-fall speed, whose dependence on radius is weaker than logarithmic.
Away from the disk and the centrifugal barrier, the gravitational potential is still 
dominated by the DM, so the accretion rate can be estimated at 
\begin{eqnarray}
\dot M(R) = \frac{M_{\rm gas}}{t_{\rm ff}}\sim \frac{M_{\rm gas}}{M_{\rm tot}}
          \frac{v_{\rm ff}^3}{G}\sim 0.4\,v_{30}^3\,M_\odot\,{\rm yr^{-1}},
\label{eq:mdot}
\end{eqnarray}
where $v_{30}\equiv v_{\rm ff}/30\,{\rm km\,s^{-1}}$ for a $2\times 10^8\,M_\odot$ DM halo, and we used the 
universal baryon fraction
for the gas-to-total mass ratio. At these radii, only the gas participates in the collapse, as the
DM is in equilibrium due to the initial conditions.

On the other hand, when the gas dominates the potential, the mass accretion rate is of order 
$\sim \alpha_{\rm v} c_{\rm s}^{3}/G \propto T^{3/2}/G$, where $c_{\rm s}$ is the gas sound speed and 
$\alpha_{\rm v}$ is the viscosity parameter in the disk \citep[e.g.][]{Shu:77,Shlosman.Begelman:89}. 
As we shall see in Section~\ref{sec:collapse}, this situation in our simulations occurs inside the 
centrifugal barrier.
For {\it local} viscosity in the disk, $\alpha_{\rm v} < 1$, but for {\it nonlocal} viscosities, 
e.g., magnetic or gravitational torques, $\alpha_{\rm v} \gtorder 1$, especially in the self-gravitating case 
where the effective $\alpha_{\rm v} \gg 1$ \citep{Shlosman.etal:90}. In this latter case, the inflow rate 
is still given by Equation~\ref{eq:mdot}, where $v_{\rm ff}$ is the virial velocity of the 
gravitational potential which is now dominated by the gas, i.e., $M_{\rm gas}/M_{\rm tot}=1$
in Eq.~\ref{eq:mdot}.
The inflow velocity will depend on the compactness of the gas distribution and can substantially exceed 
the virial speed of the DM halo.  

A high accretion rate is crucial for the formation of a SMS or disk that can develop into 
an SMBH seed. For example, \citet{Begelman:10} pointed out that an infall rate exceeding $\sim 1 \Msun$ 
yr$^{-1}$ is necessary in order to form a $10^6 \Msun$ SMS, which has a nuclear burning 
timescale of only a few million years.  
The initial SMBH seed of $100-1,000\Msun$, formed by core collapse of such a star, then grows at a highly 
super-Eddington rate inside the remaining convective envelope (the `quasistar' phase: 
\citet{Begelman.etal:08}), reaching $\sim 10^5\Msun$ or more in less than a few Myr.  

\subsection{Fragmentation of Accretion Flows}
\label{sec:fragment}

Gas fragmentation can terminate the collapse by consuming the gas supply.  In addition, the clumps formed 
during fragmentation can excite odd-mode perturbations in the gas disk, damping the bar instability which 
plays the key role in angular momentum redistribution. In order to continue collapse to very small 
spatial scales, fragmentation should be suppressed.

There are several ways to suppress gas fragmentation.  Supersonic turbulence can suppress fragmentation
via shocks that sweep the density fluctuations in a crossing time 
\citep[e.g.,][]{Padoan:95,Padoan.Nordlund:02,Krumholz.McKee:05}. 
Studies of supersonic turbulence find that the probability distribution function (PDF) of the density 
fluctuations is lognormal (i.e., Gaussian in the log) under a wide range of conditions 
\citep[e.g.,][]{Vazquez-Semadeni:94,Padoan:95,Scalo.etal:98,Ostriker.etal:99,Padoan.Nordlund:02,Krumholz.McKee:05}:
\begin{eqnarray}
p(x) = \frac{1}{(2 \pi \sigma^{2}_{\rm p})^{0.5}} \frac{1}{x} \exp [ -\frac{(\ln x -
    \overline{\ln x})^2}{2\sigma^{2}_{\rm p}}],
\label{eq:pdf}
\end{eqnarray}
where the distribution mean $\overline{\ln x} = -0.5 \sigma^{2}_{\rm p}$, $x \equiv \rho/\rho_{0}$  
($\rho_0$ being mean density), and its dispersion is $\sigma^{2}_{\rm p} \sim [\ln(1 + 3\Mach^{2}/4)]$ 
($\Mach$ is the flow Mach number). This distribution is the result of $x$ being a random variable, which 
is itself a product of independent random variables. Note that the coefficient $b=3/4$ in the definition
of $\sigma^2_{\rm p}$ is not a constant and depends on the driving mode of the turbulence. 
\citet{Federra.etal:08} found that $b$ varies between 1/3 to unity, for solenoidal and compressive driving,
correspondingly. 

The lognormal density PDF has been observed in simulations of turbulent motions on the scales of 
giant molecular clouds, where the background (unperturbed) density is uniform.  Here, we shall test the 
development of turbulence in collapsing flows with steep preexisting density gradients.

The critical overdensity required for fragmentation is $x_{\rm crit} \equiv 
(\lambda_{\rm J0}/\lambda_{\rm s})^2$, where $\lambda_{\rm J0}$ is the Jeans length at the 
mean density and $\lambda_{\rm s}$ is the turbulent length scale at which the velocity dispersion reaches 
the sound speed \citep{Padoan.Nordlund:02,Krumholz.McKee:05,Begelman.Shlosman:09}. In a supersonic 
turbulent medium, 
overdensities higher than $x_{crit}$ are unstable to fragmentation.  Estimating the mass fraction in 
Jeans-unstable fragments that collapse on a timescale shorter than the free-fall timescale as $f = 
\int^{\infty}_{x_{\rm crit}} x\frac{dp(x)}{dx} dx$, we find that $f \leq 0.02$ for $\Mach \geq 3$.  If the 
collapsing gas maintains supersonic turbulence, the amount of fragmenting gas is negligible. 

In our models, we have assumed optically thin flows. The 
spherically-symmetric, isothermal density distribution of baryons modeled  in this paper
becomes opaque at $\sim 7\times 10^{-3} X$\,pc, where $X$ is the ionization fraction. 
For temperatures encountered in this simulation, $X \ll 1$ and the electron scattering optical
depth is negligible, as is the free-free absorption.  For primordial chemical composition and the range of
temperatures $\sim 3,000-10,000$\,K, the opacity is expected to be dominated by bound-bound transitions in
hydrogen, and especially by the opacity for Lyman $\alpha$ photons. 

Radiation fields can suppress gas fragmentation by dissociating molecular gas or preventing its 
formation in the first place. If H$_2$ is absent, the gas temperature will be maintained at $> 8,000$\,K 
\citep[e.g.,][]{Omukai:01,Shang.etal:10}. The prime agent of H$_2$ dissociation can be an externally-produced 
Lyman-Werner ($\sim 11$\,eV$-13$\,eV) 
continuum, e.g., from neighboring Pop\,III stars. \citep[e.g.,][]{Dijkstra.etal:08}.  H$_2$ formation could 
be inhibited by the high temperatures likely to obtain in the collapsing gas.  The cooling efficiency of  
Ly$\alpha$ produced in the accretion flow at $T \sim 10^4 \K$ is limited by the high optical depths. For a 
sufficiently 
high column density of neutral hydrogen, collisional de-excitation can decrease the escape fraction of 
Ly$\alpha$, preventing the gas from cooling below $\sim 8,000$K and suppressing H$_2$ formation thermally 
\citep[e.g.,][]{Spaans.Silk:06,Schleicher.etal:10,Latif.etal:11}.

One might speculate whether Ly{$\alpha$} photons generated within the accretion flow could be upscattered 
into the Lyman-Werner continuum before escaping, thus producing a dissociating flux self-consistently.  
For an isothermal density profile, we estimate the required Ly$\alpha$ optical depth (measured in the 
Doppler core) of $\tau_\alpha \gtorder 10^{14}$ \citep[e.g.,][]{Harrington:73,Neufeld:90}, which can be 
attained at radii $\ltorder 10^{-3}$ pc for the parameters of our models.  Other processes, however, are 
likely to allow the 
energy in these upscattered Ly$\alpha$ photons to leak away before they can escape.  For one thing, these 
photons can be absorbed by even small amounts of dust. Extrapolating Figure\,18 of \citet[][]{Neufeld:90} 
to our column densities (which can be inferred from the next section), we find that a metallicity as low as 
$10^{-8}$ solar will do the job.  An additional process that can destroy the Ly$\alpha$ photons is the 
resonant pumping of H$_2$, followed by fluorescent  decay to vibrational levels.  

We note that fragmentation might be avoidable in the presence of supersonic turbulence, even if H$_2$ is 
present \citep[e.g.,][]{Begelman.Shlosman:09}. This happens because the fraction of fragmenting gas 
decreases with $\Mach$.

In our simulations, we assume that H$_2$ is destroyed and do not analyze its contribution to the gas cooling. 
We also neglect magnetic fields and their effects on the turbulent flow, and model 
gravitational collapse within the DM halo hydrodynamically. This is justified because $B$-fields are 
expected to be weak at high $z$, and because of the high temperature of the flow, $\gtorder 1,000$\,K.  
At these high temperatures, the value of the critical $B$-field strength needed to limit the compression 
in supersonic isothermal shocks is also higher \citep{Padoan.etal:07}. 

\section{Numerical technique}                    %
\label{sec:method}                               %

\subsection{The numerical code ENZO}
\label{sec:enzo}

In order to test the theoretical arguments made in Section~\ref{sec:theory}, we use an Eulerian adaptive 
mesh refinement (AMR) code ENZO-2.1, which has been tested extensively and is publicly available 
\citep{Bryan.Norman:97,Norman.Bryan:99}.
ENZO uses a particle-mesh $N$-body method to calculate the gravitational dynamics including collisionless 
DM particles, and a second-order piecewise parabolic method \citep[PPM,][]{Bryan.etal:95} to solve 
hydrodynamics. The structured AMR used in ENZO places no fundamental restrictions on the number of 
rectangular grids used to cover some region of space at a given level of refinement, or on the number of 
levels of refinement \citep{Berger.Colella:89}. A region of the 
simulation grid is refined by a factor of 2 in length scale if the gas density is greater than $\rho_0 N^l$, 
where $\rho_0$ is the minimum density above which refinement occurs, $N = 2$ is the refinement factor and 
$l=25$ is the maximal AMR refinement level. This refinement corresponds to a spatial resolution of 
$\sim 10$\,AU. 

The \citet{Truelove.etal:97} requirement for resolution of the Jeans length, i.e., at least 4 cells, has been 
verified. However, the actual resolution of the Jeans length in our simulations exceeds this criterion, depending
on the distance from the center, because of the baryon density dependence. Specifically, throughout most of the 
spherical collapse region, $10-10^3$\,pc, the Jeans length is resolved with $\gtorder 100$ cells. Between 1 -- 
10\,pc, it is resolved by $\sim 32$ cells, for 0.001 -- 1\,pc by 10-30 cells, and inside 0.001\,pc by 4 cells.
This estimate is based on the baryon properties only, and ignores the effect of DM.
If the latter is taken into account, the Jeans length increases by a factor of a few.
Therefore, our resolution of the Jeans length increases correspondingly.

We have also run test models 
resolving the Jeans length with 64 cells, as required by e.g., \citet{Sur.etal:10}, \citet{Federrath.etal:11}, 
\citet{Turk.etal:12} and \citet{Latif.etal:13}  in the MHD simulations.
No qualitative diference in the evolution has been found, except a slight, $\sim 10\%$, increase in the
onset time of the 2nd stage of the gravitational collapse, i.e., of the central runaway 
(see Section~\ref{sec:simresult} for the definition).

ENZO follows the non-equilibrium evolution of six species: $\rm H, \; H^{+}, \; He, \; He^{+},  \; He^{++}$, 
and $e^{-}$ \citep{Abell.etal:97,Anninos.etal:97} in a gas with primordial composition.  It calculates 
radiative heating and cooling following atomic line excitation, recombination, collisional excitation and 
free-free transitions.  Radiative losses from atomic cooling are computed in the optically-thin limit.  As 
we discussed in Section~\ref{sec:theory}, there are several radiation transfer processes that have been 
suggested to prevent $\rm H_2$ formation. In order to include these effects without implementing a full 
radiative transfer calculation, we exclude the chemistry and cooling related to $\rm H_2$ in this paper.

\subsection{Simulation Setups}
\label{sec:IC}

For a gas with primordial composition and no $\rm H_2$, the earliest collapse can occur in DM halos whose 
virial temperatures exceed $\sim 10^4\,\K$. For this reason we set up a spherical DM halo with 
$M_{\rm vir}\sim 2 \times 10^8h^{-1}\,\Msun$ at $z=15$. According to the top-hat model, such a halo will have 
a virial radius of $R_{\rm vir} \sim 945h^{-1}\,\rm pc$ and a virial temperature $T_{\rm vir} \sim 32,000\,\K$, 
according to the WMAP5 cosmology ($\Omega_{\rm m} = 0.279$, $\Omega_{\rm b} = 0.0445$, and $h=0.701$) 
\citep{Komatsu.etal:09}. We simulate this DM halo within a $(6\,\rm {kpc})^3$ computational domain.

For the DM halo, we assume an isothermal sphere.  This halo model is similar to the universal DM halo model 
\citep[NFW,][]{NFW:96,NFW:97} at the spatial scales of interest. The isothermal model also provides a simple 
analytical form for the velocity distribution function which allows us to generate a live isotropic DM particle 
distribution that maintains a stable equilibrium with $10^6$ particles. 
We introduce a DM core radius, $R_{\rm dm}$, within which the DM density is approximately 
constant \citep[e.g.,][]{Elzant.etal:01,Tonini.etal:06,Romano.etal:08,Primack:09}. Therefore, $R_{\rm dm}$ 
plays the role of the King radius for a nonsingular isothermal sphere. We vary 
$R_{\rm dm}$ to see the effect of DM halo structure on the 
development of gas collapse. In particular, we implement a 
number of simulations, models A -- E, with different DM density core sizes, as given in Table~\ref{tab:table1}.  

The DM halo and the gas have been laid down at virial and pressure (for the gas) equilibrium, so no
transient adjustment occurs. The gas halo has a $R_{\rm core} = 100\,\rm pc$ 
constant density core at the start of the simulations. The gas, however, instantly cools down from the 
virial temperature to about 10,000\,K, and so its pressure support is negligible.
The total gas mass in the halo is estimated from the 
cosmological baryon fraction, $\Omega_{\rm b}/\Omega_{\rm m} \sim 0.16$. The angular momentum of the halo gas 
is computed from $J = \lambda \sqrt{2} M_{\rm vir} v_{\rm c} R_{\rm vir}$, defined in Section~\ref{sec:theory}. 
In large-scale $N$-body cosmological simulations, \citet{Bullock.etal:01} has found that the specific angular 
momentum in DM halos roughly follows $j(R) \sim R^{1.1}$, close to a flat rotation curve. We therefore impose 
a constant rotation velocity on the gas outside the core, assuming that the rotational axis runs parallel to 
the $z$-axis. For $R\ltorder R_{\rm core}$, we assume solid body rotation for the gas, i.e., 
$j(R)\sim R^2$  \citep[e.g.,][]{Samland.Gerhard:03,Heller.etal:07}.
The outer boundary of the model we smooth the sharp density cutoff at the 
virial radius by adopting the gas density profile $\sim R^{-3}$ beyond $R_{\rm vir}$.  

We also ran a number of models with modified initial conditions. One such representative model is discussed as 
model Dmod, i.e., modified model D, in Section~\ref{sec:modelDmod}. In this model we randomized the 
orientation of the initial angular momentum in spherical shells, keeping the mean-square angular momentum 
unchanged. For this to happen, we have slightly decreased $j$ of the inner shells and compensated the outer
shells by a slight increase in $j$. This was done in order to mimic 
cosmological initial conditions, based on the simulations by \citet{Romano.etal:09}, in particular on their 
Figure~19. 

To verify that the DM halo is indeed formed in equilibrium, we tested DM-only models and confirmed their 
stability, especially in the central region. Additional tests have been run for model B with an isothermal 
equation of state for the gas; these show very similar evolution. We also tested a model with 
an adiabatic equation of state --- the gravitational collapse did not proceed far in this model, as expected.  
 
\begin{table}
\caption{\bf DM Core Radius and Central Runaway Collapse Time}.  
\centering
\begin{tabular}{lcccccccc}
\hline
 Models \# & DM Core Radius $R_{\rm dm}$ (pc)& Collapse Time (Myr)\\
\hline
 A   &  0.10& 2.1 \\
 B   &  0.40& 4.7 \\
 C   &  0.75& 8.7 \\
 D   &  1.50& no collapse \\
 E   &  6.00& no collapse \\
\hline
 Dmod&  1.50& 4.5 \\
\hline
\end{tabular}
\tablecomments{The collapse time given in the third column is time between the start of the simulation and
the start of the central runaway collapse.}
\label{tab:table1}
\end{table}

\begin{figure*}
\centerline{
  \includegraphics[width=0.5\textwidth,angle=0] {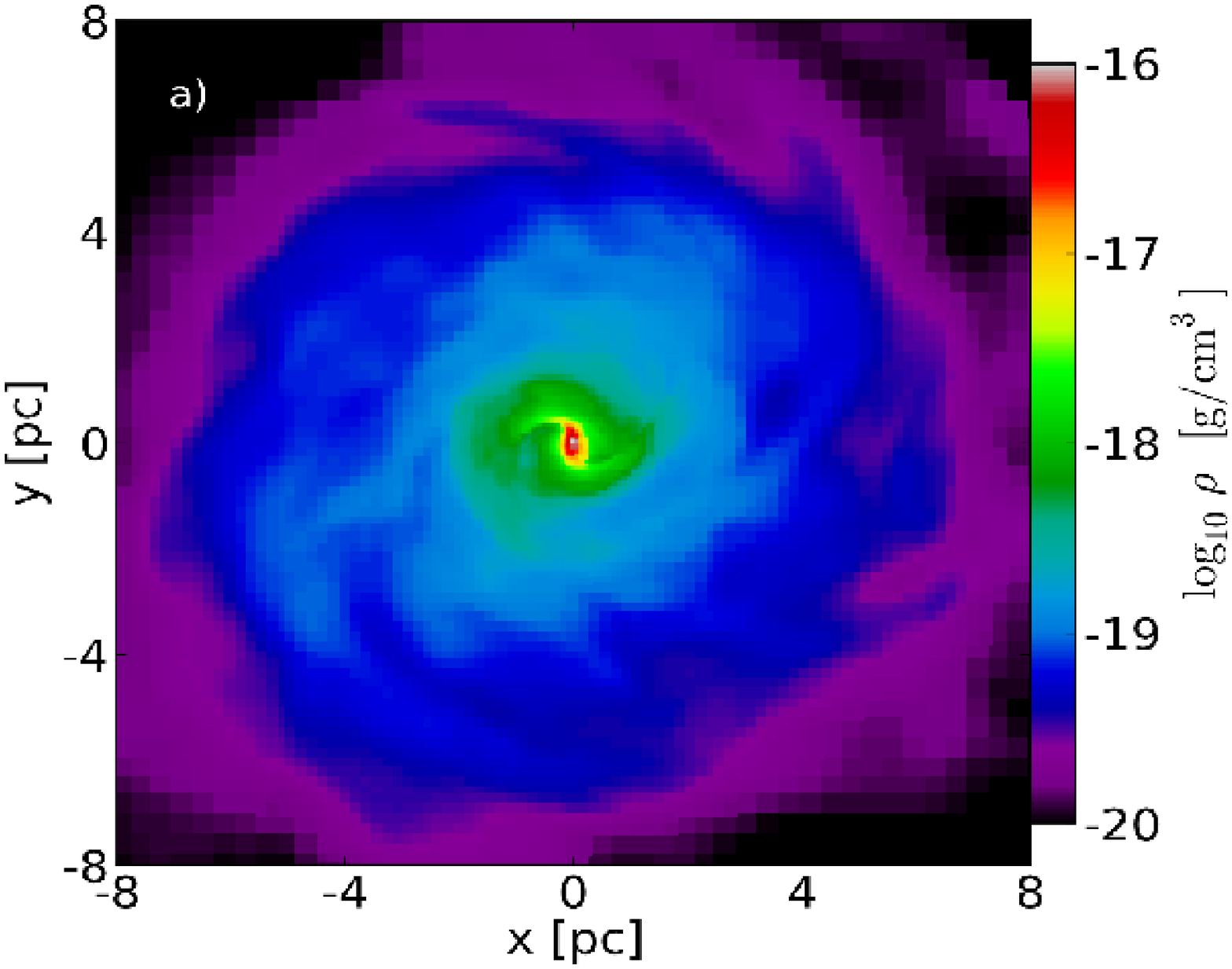}
  \includegraphics[width=0.5\textwidth,angle=0] {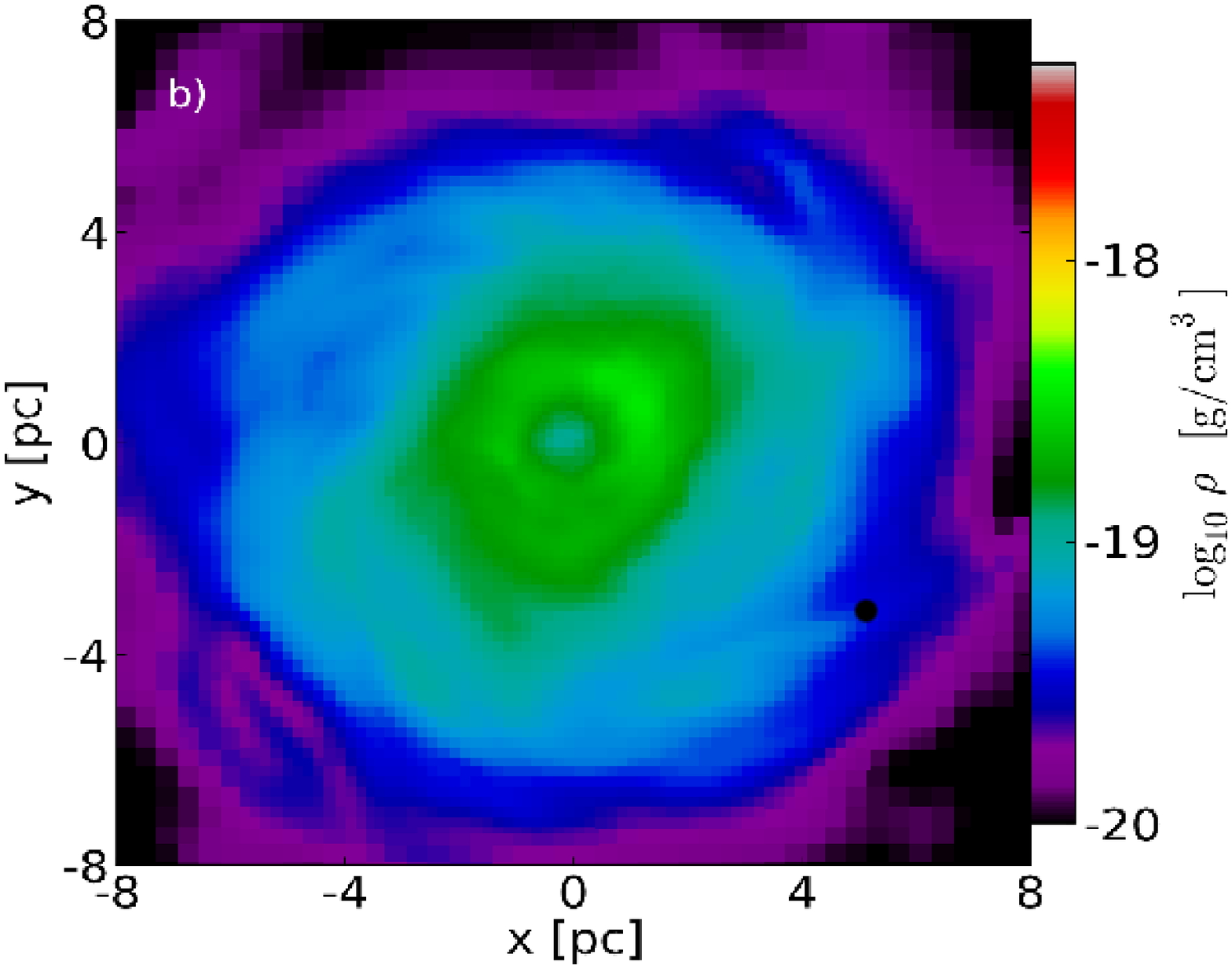}
}
\caption{
Density-weighted projection of density (see text) of {\it (a)} model B, and {\it (b)} model D, with thickness 
$\Delta z = 2$\,pc about 
the equatorial plane, at $t=4.6$\,Myr.  These face-on views show gas disks that have formed from the collapsing 
gas, immediately after the gas disk in model B  has undergone runaway collapse at its center. Model B shows the 
formation of a gas bar that redistributes angular momentum and drives a strong mass inflow, whereas the gas disk 
in Run D maintains a steady density profile. Both gas disks are intermittently turbulent.  Snapshot resolutions 
are 0.01\,pc (left) and 0.1\,pc (right). Box size is 16\,pc. 
}
\label{fig:frame4.6}
\end{figure*}

\section{Results}                 %
\label{sec:simresult}             %
 
We would like to emphasize several important points about the gravitational collapse modeled here, before we 
show the results of numerical simulations. First, the evolution described here truly proceeds from inside
out. Development of the central mass accumulation and, therefore, of the prospective seed SMBH, happens on timescales
much shorter, $\ltorder 10$\,Myr, than the free-fall timescale for the DM halo, $\sim 80$\,Myr. Hence, the dynamics
of the outer parts on scales of $\sim 0.1-1$\,kpc, although interesting in itself, appears to be irrelevant for 
the central regions of $\ltorder 10-50$\,pc, which dominate the formation of the SMBH seed. Next,
the gravitational collapse proceeds in two stages. The first stage involves infall, braking and the formation 
of a gaseous disk-like configuration inside the centrifugal barrier. All the models exhibit this phase.  
Models A -- C show the second stage of gravitational collapse in which the inner part of the disk, 
$r\ltorder 1-5$\,pc, develops a runaway which proceeds to the point where the numerical evolution has been 
terminated, at $\sim 10^{-4}$\,pc $\sim 20$\,AU. Our task will be to explain this evolution. We shall 
emphasize models B and D as representative of 2-stage and 1-stage collapse, respectively. Model Dmod is 
discussed separately.

\vspace{0.5cm}
\subsection{Loss of axial symmetry and central runaway}
\label{sec:collapse}

As the gas has little rotational support, it goes into nearly free-fall collapse, which develops from inside 
out because of the shorter dynamical timescales at small $r$ and larger gravitational accelerations there.
This leads to the establishment of $\rho\sim R^{-2}$ density profile and to a largely homologous collapse, except 
when and where the angular momentum becomes important.
There is little difference among models at this stage, 
as they differ only in the value of the DM core radius $R_{\rm dm}$, and, therefore, in the depth of the DM 
potential well. Figure~\ref{fig:frame4.6} shows face-on density-weighted projections of the 
density\footnote{$\Sigma_{\rm i}(\rho_{\rm i} W_{\rm i})/\Sigma_{\rm i}(W_{\rm i})$, where the weight function 
$W_{\rm i}$ is $\rho$} of the gas disk at $t=4.6$\,Myr
in models B and D, with $R_{\rm dm}\sim 0.4$\,pc and 1.5\,pc, respectively. Unless mentioned, other models behave 
similarly.

\begin{figure}
\centerline{
  \includegraphics[width=1.15\columnwidth,angle=0] {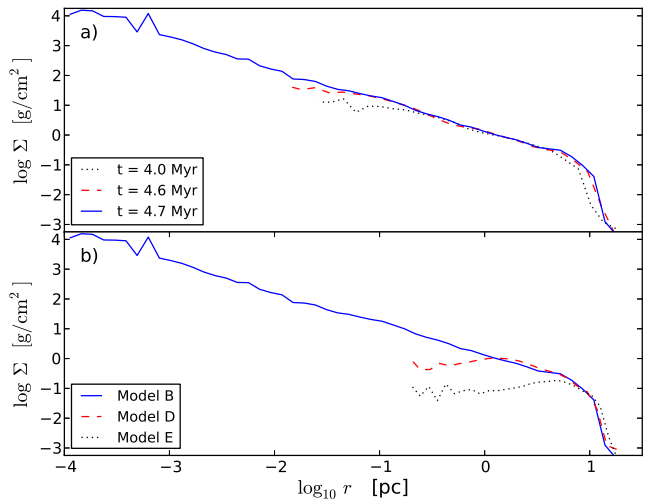}
}
\caption{Surface density of the growing disk, averaged over annuli, during the central runaway. 
{\it (a)} model B (solid blue) at 
$t\sim 4.0-4.7$\,Myr. {\it (b)} models B (solid blue), D (dashed red) and E (dotted black) are 
compared at $t=4.7$\,Myr. 
}
\label{fig:SurfDens}
\end{figure}
\begin{figure}
\centerline{
  \includegraphics[width=1.15\columnwidth,angle=0] {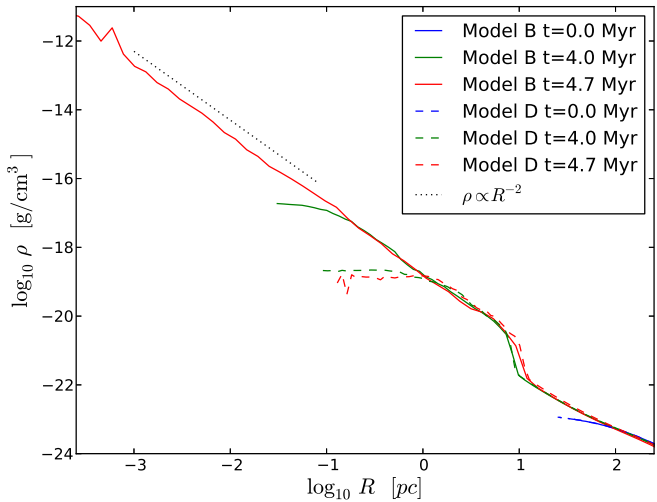}
}
\caption{
Evolution of gas volume density profiles averaged over spherical shells, for models B and D. Both simulations 
start from identical initial 
conditions. Owing to the deeper DM potential, the gas disk in model B has a steeper mass distribution in 
the center (see the profiles at $t=4.0$\,Myr).  At $t=4.7$\,Myr, model B shows runaway collapse (strong 
mass inflow) driven by gaseous bars, while the model D disk shows marginal evolution from $t=4.0$\,Myr to 
$t=4.7$\,Myr.  The density profile of the runaway collapsing gas structure is close to that of a singular 
isothermal sphere (dashed line).  
}
\label{fig:profile}
\end{figure}

\begin{figure*}
\centerline{
   \includegraphics[width=2.1\columnwidth,angle=0] {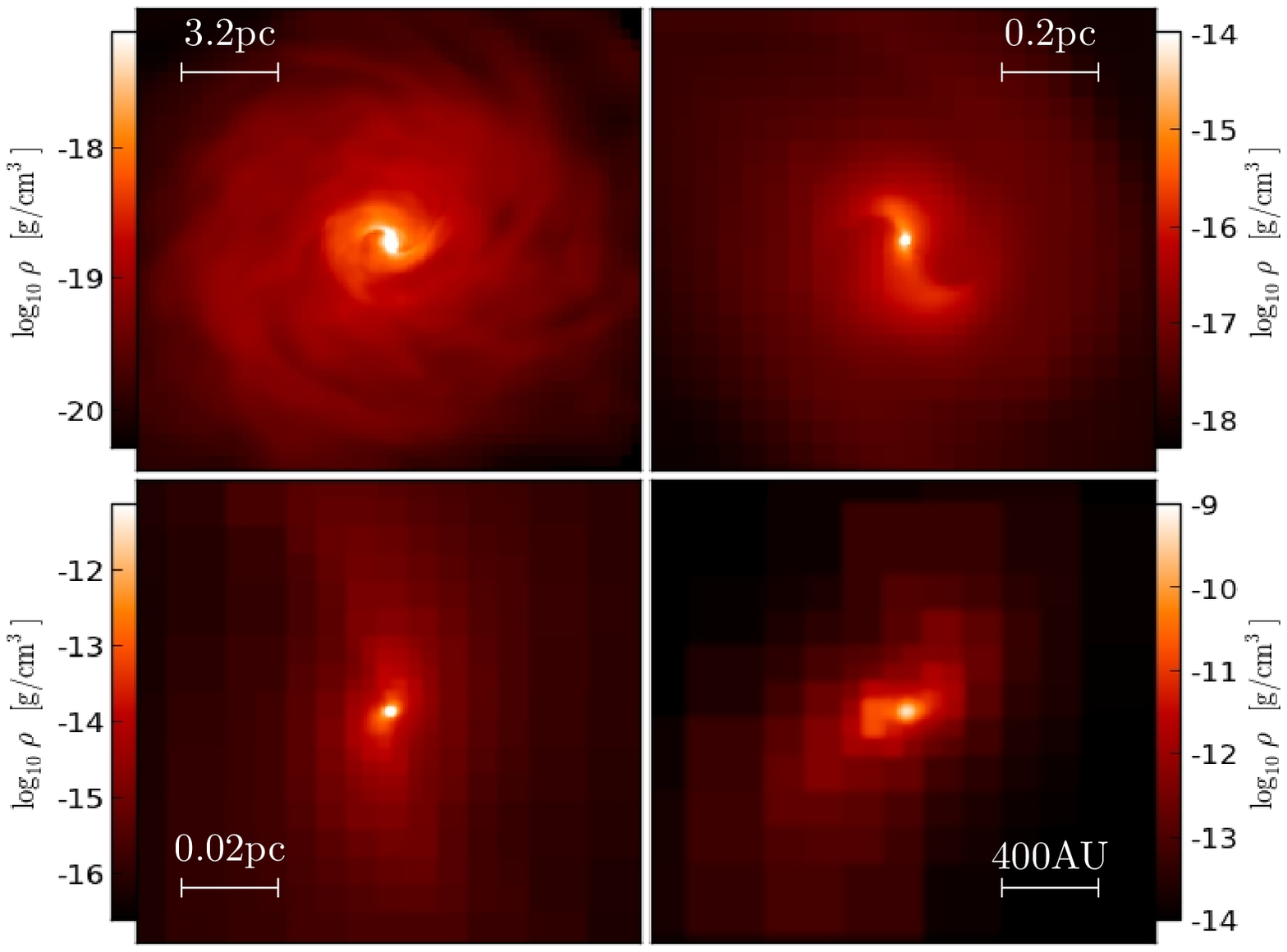}
}
\caption{Face-on density-weighted projection of density showing the gaseous bar cascade in model B on different 
spatial scales. {\it Upper left:} 16\,pc box at $t=4.6$\,Myr;   
{\it Upper right:} 1\,pc; {\it Lower left:} 0.1\,pc; {\it Lower right:} 0.01\,pc at $t=4.7$\,Myr, the time of the 
runaway collapse at the center. The slice thickness $\Delta z$ is (a) 2\,pc, (b) 0.5\,pc, (c) 0.1\,pc and 
(d) 0.01\,pc.  
}
\label{fig:slice4.6}
\end{figure*}

During the first stage of collapse, the baryon angular momentum flows inward.  The inner gas is the first to 
reach the centrifugal barrier, and develops a disk-like configuration at $r\sim 1$\,pc. The disk grows quickly 
in radius and mass, and by $t\sim 4$\,Myr, the snapshots display well-developed gas disks with radii $\sim 
10$\,pc. The disk boundaries, both in $r$ and in $z$, are delineated by standing shocks, discussed later. The 
surface densities of all models are well-approximated by a power-law $\Sigma\sim r^{-n}$ with $n\sim 1.3$.  
Figure~\ref{fig:SurfDens} displays the evolution of the surface density for model B, as well as comparisons 
of the surface densities at $t = 4.7$ Myr among models B, D and E.  The disks are truncated at $r\sim 10$\,pc, 
as clearly seen in this figure.  The outer surface density profiles of all disks, in models A to E, are very 
similar, but in the central $\sim 1-2$\,pc the density profiles differ: the density in model B within the 
central 0.1\,pc is more than an order of magnitude higher than the density in model D.

\begin{figure}
\centerline{
  \includegraphics[width=1.17\columnwidth,angle=0] {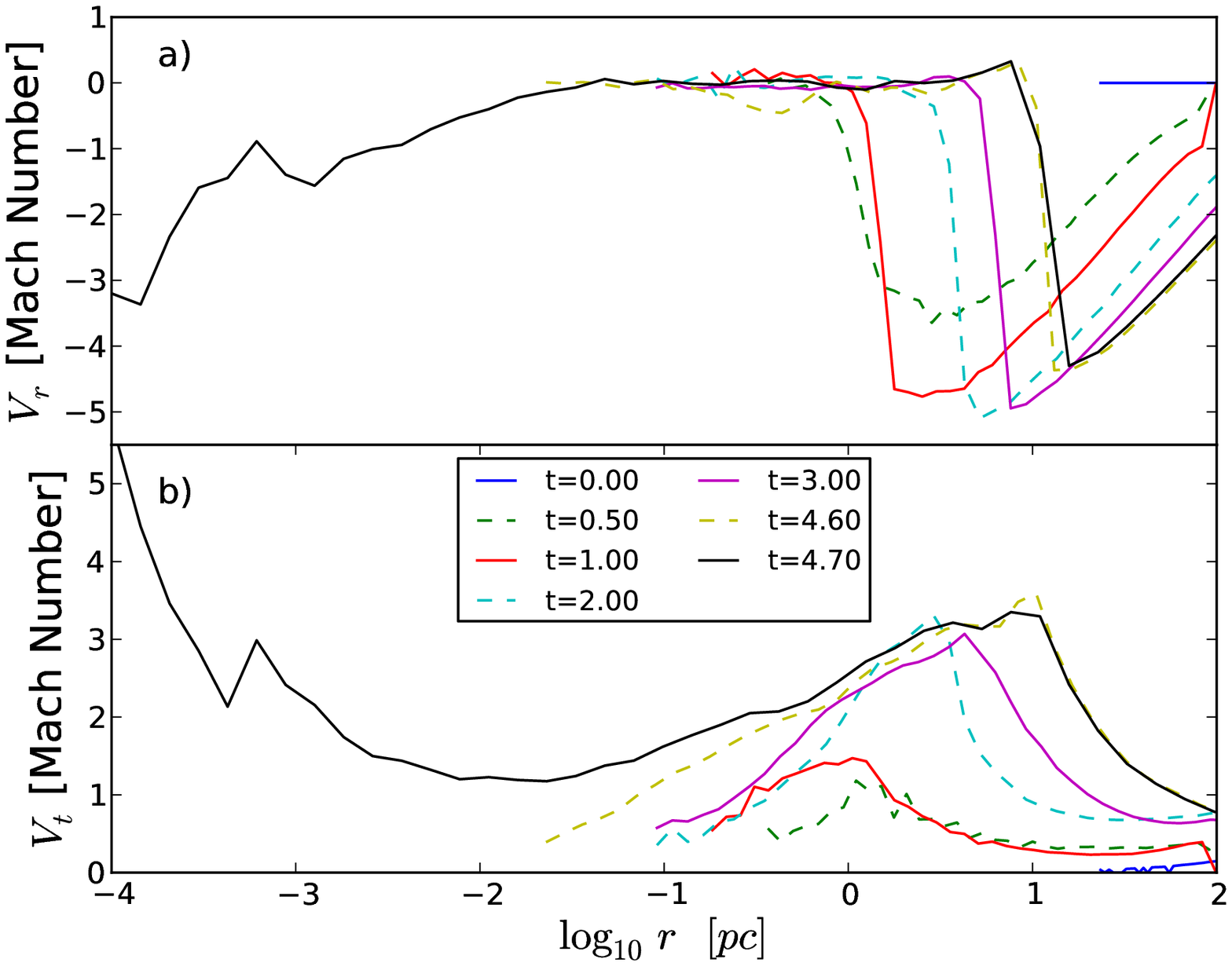}
}
\caption{{\it (a)} Evolution of the radial velocity profile as a function of $r$ in the disk plane for 
model B, normalized by the sound speed (i.e., in Mach numbers). {\it (b)} Evolution of the azimuth-averaged 
tangential velocity (in Mach numbers) as a function of $r$ in the disk plane for model B.  
}
\label{fig:VrMach}
\end{figure}

\begin{figure}
\centerline{
  \includegraphics[width=1.15\columnwidth,angle=0] {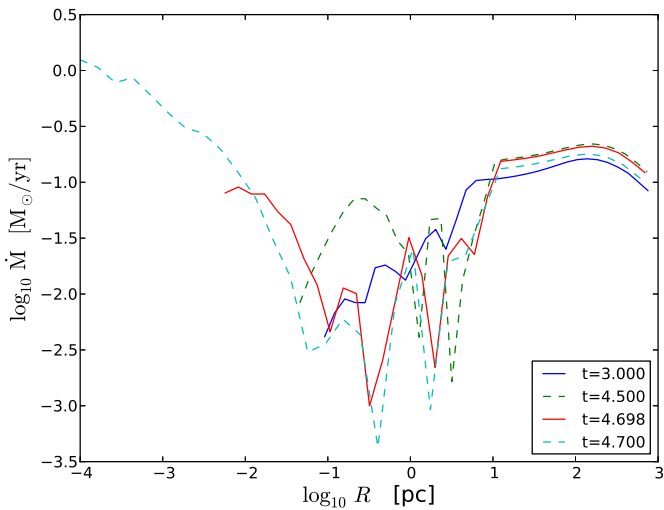}
}
\caption{Evolution of the mass accretion rate in model B at various times, from before the second stage of 
collapse at $t = 3$\,Myr, through the runaway at $t\sim 4.7$\,Myr.
}
\label{fig:Mdot}
\end{figure}

\begin{figure}
\centerline{
  \includegraphics[width=1.15\columnwidth,angle=0] {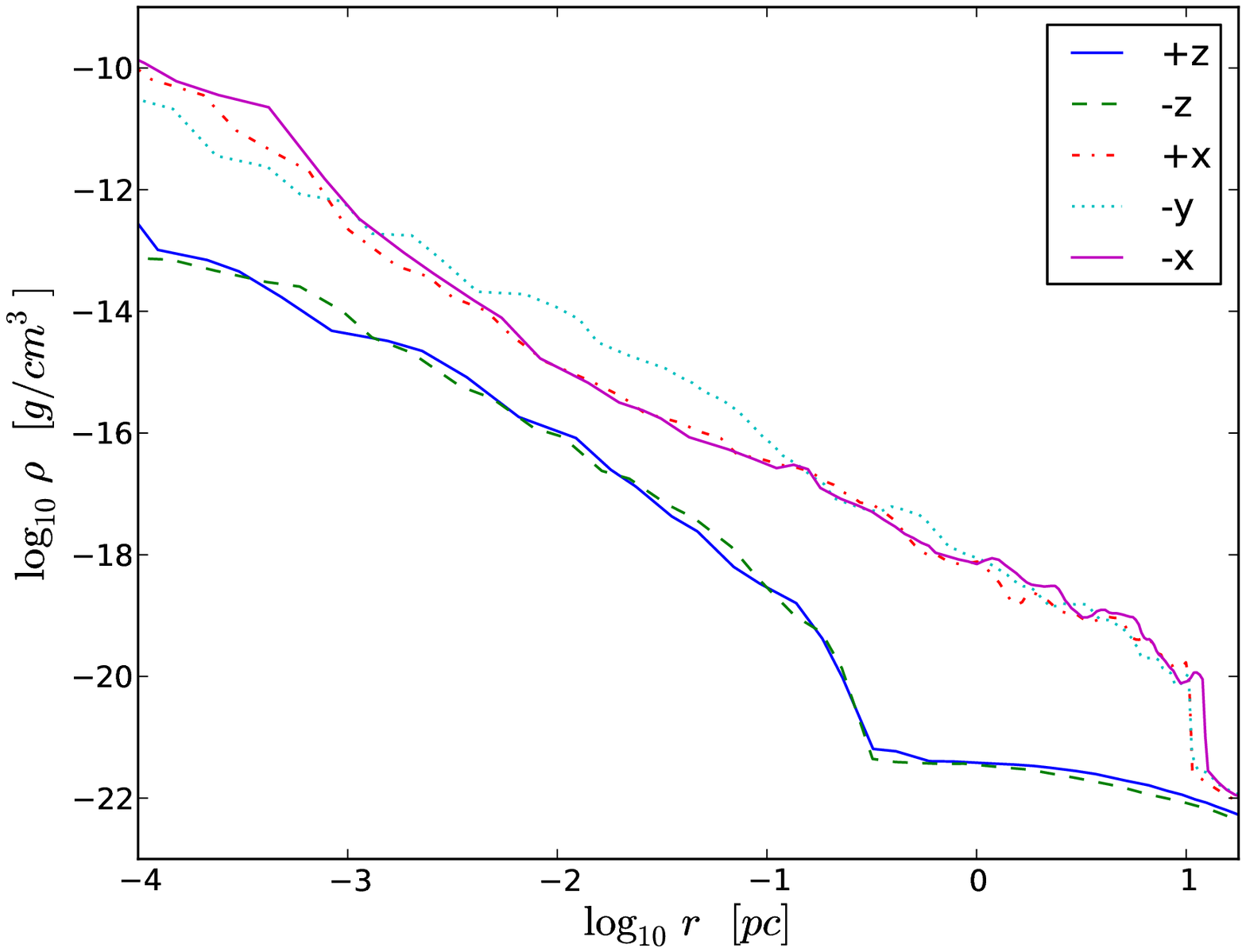}
}
\caption{Density profiles along the axes noted at the runaway collapse time $t\sim 4.7$\,Myr for model B. The 
profile along the $z$-axis has been measured at $x=y=0$. Note the positions of the radial shock at $r\sim 10$\,pc 
and the surface shock at $z\sim 0.5$\,pc.
}
\label{fig:EllipDen1}
\end{figure}

The DM density profiles show little evolution even in the central regions. In models B, D and E 
(Figure~\ref{fig:SurfDens}), and also in all other models, we observe the trend that more cuspy DM halos develop 
higher surface and volume density gas disks. The disk in D has a surface density about a factor of $\sim 10$
times higher than E. The disk volume density also appears higher in more cuspy DM 
halos (Figure~\ref{fig:profile}). This difference is clearly observed within the DM core radius $R_{\rm dm}$ 
of less cuspy halos.

Over the next time period of a few$\times 10^5$\,yr, the disk surface density grows as a result of accretion. 
The second stage of the collapse is reached only in models A -- C, and is characterized by both baryonic 
angular momentum outflow and continuing mass influx. On larger scales, just outside the centrifugal barrier, $j$ 
increases, moving the barrier outward.  Inside the centrifugal barrier, the disk develops a global instability, 
dominated by the Fourier component $m=2$ --- a bar-like mode, which transfers $j$ to the outer gas and to the DM.
For example, model B exhibits a well defined gaseous bar in the central $\sim 1-2$\,pc of the disk at 
$\sim 4.7$\,Myr. The appearance of this bar-like mode is characteristic also of smaller spatial scales of this 
model (Fig.~\ref{fig:slice4.6}), as we shall analyze below.
The timescale for the onset of this stage of the central runaway is given in 
Table~\ref{tab:table1}, and is increasing along the sequence A\,$\rightarrow$\,C, i.e., toward less cuspy halos.  
On the contrary, models D and E display a weak central depression which has the appearance of a ring-like 
structure around the center and no $m=2$ axial symmetry-breaking. 
At a later stage, these models exhibit off-center fragmentation.

Figure~\ref{fig:frame4.6}a  confirms the crucial 
role of gravitational torques in shaping the inner mass distribution via angular momentum transfer, which 
requires a substantial departure from axial symmetry. This symmetry breaking gives rise 
to the lowest modes, $m = 1$, 2, and occasionally $m=3$. Only $m=1$ exhibits a mild displacement of the center
of mass of the disk, while $m=2$ has the appearance of a strong gaseous bar which 
drives spiral structure, and $m=3$ shows up as tri-armed spirals. We follow the runaway collapse to a spatial 
scale of
$\sim 10^{-4}$\,pc (20 AU), and stop the calculation there because the flow is expected to become opaque and
enter a dynamical regime where radiation pressure must be taken into account (see Section~\ref{sec:fragment}).  

Owing to the AMR nature of ENZO, the small scale structure in Figures~\ref{fig:SurfDens} and \ref{fig:profile} 
is only resolved when a given region exceeds the density threshold for a new refinement.  The density profile in 
model B is resolved to $\sim 20$\,AU at the end of the simulation, while model D gas only reaches a resolution 
of $\sim$ 0.1 pc. Resolving the $\sim 20$\,AU scale means that the gas inflow has reached this scale by 
overcoming the angular momentum barrier. It demonstrates that the gravitational torques resulting from axial 
symmetry-breaking at $\sim 1$\,pc in model B trigger continuous gas inflow on progressively smaller scales. 
Finally, Figure~\ref{fig:profile} demonstrates that the timescale of this inflow is very short, from 
$t=4.0$\,Myr to $t=4.7$\,Myr, a truly runaway collapse with a dynamical time of $\ltorder 10^6$\,yrs.

For the idealized case of spherical accretion within an external gravitational potential given by 
$\rho \propto R^{-2}$, the gas mass flux is given by $\dot M\sim R^2 \rho v_{\rm R}$ 
with $v_{\rm R}$ weakly dependent on $R$.
Figure~\ref{fig:profile} confirms that the density profile of the collapsing 
gas in model B tends to the isothermal density profile, $\rho\propto R^{-2}$, for $R\ltorder 10$\,pc. 
It decreases sharply outside this radius due to the shock at the centrifugal barrier, then continues
with the same slope at larger radii. (Note that all models start with 
a flat gas density core of 100\,pc.) Figure~\ref{fig:VrMach}a shows 
the radial profile of the inflow velocity 
measured in the disk plane. We can divide the $r$-range into the collapsing gas in the region outside 10\,pc, 
the disk plateau at $r\sim 1-10$\,pc, and the inner region of runaway collapse --- all at the 
time of the central runaway, $t\sim 4.7$\,Myr. 
The stronger than expected increase of Mach number with decreasing radius has its origin in a 
combination of temperature decrease (due to an increased gas density) and the steepening increase of
$v_{\rm R}$ toward the center, as the result of the runaway collapse of a finite gas mass.

Thus the outermost region is dominated by free-fall kinematics because the angular momentum is low. The density 
profile is maintained as $\rho\sim R^{-2}$ if one neglects the weak variation of $v_{\rm R}$. 
The intermediate disky region has $v_{\rm R}\sim const.$ and the same $-2$ slope of log\,$\rho$. 
The innermost collapse exhibits a rapidly increasing and steepening infall velocity 
(Fig.~\ref{fig:VrMach}a,b). 
The accretion flow is mostly rotationally supported in the disky 
region of $1-10$\,pc, as well as in the central region, where the self-gravitating collapse proceeds with a
non-negligible angular momentum.
The time evolution of $v_{\rm R}$ in the central region reflects the self-gravitating collapse developing
there, and a dynamical decoupling of the gas from the DM background potential.

The measured mass accretion rate, $\dot M(R)$, appears to be approximately flat, $\sim 0.1-0.2\,M_\odot\,{\rm 
yr^{-1}}$, exterior to the radial shock, in agreement with Eq.~\ref{eq:mdot} (Fig.~\ref{fig:Mdot}). In the disk 
region, $\dot M(R)$ decreases abruptly to $\sim 0.01\,M_\odot\,{\rm yr ^{-1}}$ at $t\sim 3$\,Myr, and drops by an
additional order of magnitude during the second stage of the collapse, when the material is drained by
the central runaway. At this latter time, $t = 4.7$\,Myr, the 
central runaway is accompanied by peak values of $\dot M(R)\sim 2\,M_\odot\,{\rm yr^{-1}}$. The decrease in the 
accretion rate in the range of $r\sim 1-5$\,pc is a reflection of mass conservation, as the bar-like mode 
channels gas inward at a higher rate than it can be resupplied by disk accretion, and thus drains this region. 
This minimum in $\dot M(R)$  is found in the region where the turbulence induced by the radial shock decays.
In Section~\ref{sec:MinMass}, we use this radius in model B to estimate the baryon mass that participates in the 
second stage of the collapse, and therefore, the mass of the seed SMBH.

The high mass accretion rate is an important ingredient of early SMBH formation in the direct collapse model
\citep[e.g.,][]{Begelman.etal:06,Begelman.etal:08,Begelman.Shlosman:09,Begelman:10}.
The central runaway region exceeds the mass accretion rate given in Eq.~\ref{eq:mdot} by about an order of 
magnitude. The reason for this lies in the fact that during the second stage of the collapse, the inflow 
velocities are determined by the compactness of the gas distribution (Section~\ref{sec:AngMom}) and not by the
DM halo virial velocity. Under these conditions, the ratio $M_{\rm gas}/M_{\rm tot}\sim 1$ in Eq.~\ref{eq:mdot},
and the free-fall velocity, $v_{\rm ff}$, becomes a rapidly growing function of time.
As a result, the characteristic
$r$-profile of radial infall velocity during this stage is characteristic of the `avalanche' behavior of a 
dynamically decoupled gas. This radial velocity is expected to grow sharply with time. The timescale of
this process depends only on the mass involved in the collapse, and on the ability of the gas to cool ahead
of the free-fall time. If the latter condition is fulfilled, the gas will collapse to an infinite density
in a finite time.

We note that the gas joining the disk experiences strong radial and 
vertical ($z$-axis) shocks, as is seen in Figure~\ref{fig:VrMach}a (and Figs.~\ref{fig:EllipDen1} 
and \ref{fig:outerTurb}). But there is no associated shock in the azimuthal motion. The face-on and edge-on 
velocity fields on scales of $\sim 10$\,pc are shown in Figure~\ref{fig:outerTurb}. While the face-on view is 
dominated by rotation, the edge-on view exhibits turbulent motions dominated by vortices, as we discuss in 
the next section.

We next examine the geometry of the collapsing gas during the runaway stage: is it best approximated as
one-dimensional  collapse with a spherical symmetry, where the angular momentum is completely unimportant, or 
does it exhibits a cylindrical or disk-like geometry?  To answer this, we compiled the density profiles of the 
collapsing gas in model B at $t\sim 4.7$\,Myr, along two directions in the equatorial plane, $\pm x$ and  
$\pm y$, and along the $\pm z$-axis (Fig.~\ref{fig:EllipDen1}). We verify that the collapse outside the 
centrifugal barrier proceeds in a spherical fashion, which is understandable because the angular momentum in
this region is not important. At $t\sim 4.7$\,Myr, the radial shock in the equatorial plane is positioned at
$r\sim 10$\,pc (see also Figure~\ref{fig:VrMach}a). 

Inside the radial shock marking the centrifugal barrier, i.e., between $r\sim 1$\,pc and 10\,pc, the equatorial
density increases substantially while the density along the $z-$axis is almost constant.  At $\sim 1$\,pc, the
density along the $z$-axis is $\sim 10^{-3}$ times the density measured along the $x$- or $y$-axis at the 
same radial distance.  The reason for this abrupt change in density ratio lies in the relative positions of the
radial shock and the surface shock (i.e., the shock in the vertical velocity, at roughly constant $z$) in the 
disk. While the centrifugal barrier stops the cold inflow toward the 
rotation axis at $r\sim 10$\,pc, the position of the disk surface shock, i.e., the disk thickness, is 
determined by the postshock temperature which never exceeds $\sim 10^4$\,K, as seen in Fig.~\ref{fig:Temp_rho}. 
 
A strong surface shock is maintained by the infall along the $z$-axis at $\sim 0.5$\,pc (Fig.~\ref{fig:EllipDen1}).
Both radial and surface shocks are slowly moving outward with time. Most interestingly, at the time of the central 
runaway, the surface shock collapses toward the equatorial plane around $x=y=0$. 
The `peanut' shape of the surface shock at this time can be inferred from Fig.~\ref{fig:outerTurb}. 
Figure~\ref{fig:EllipDen1} reveals that within the central region 
the second stage of the collapse proceeds in the self-similar fashion, preserving the disk-like configuration.
We have tested this by plotting these distributions at various times.
But even as a single snapshot, Figure~\ref{fig:EllipDen1} nevertheless reflects the evolutionary trend 
because different $r$ 
are associated with different dynamical timescales. Hence the fact that $\rho(z)/\rho(r)\sim const.$ for
different $r$ means that this ratio also does not evolve in time --- a clear sign of self-similarity
during the central runaway. Hence the gas configuration remains dynamically cold, 
sustaining the bar cascade that efficiently transfers angular momentum outward (Fig.~\ref{fig:slice4.6}).  

We now return to Figure~\ref{fig:slice4.6}, which displays the non-axisymmetric bar-like perturbations on a 
number of spatial scales. We first observe this instability developing on scales $\sim 1-2$\,pc, and then 
propagating inward. The dynamics of
gaseous bars has been analyzed by Englmaier \& Shlosman (2004), who focused on the decoupling of a gaseous bar 
from large-scale stellar bars. This decoupling is associated with the rapidly increasing pattern speed of the 
gaseous bar and its radial contraction.

We measured the pattern speed of the gaseous bar in model B on scales of 1\,pc and and 0.01\,pc. The resulting 
values of the pattern speeds, $\Omega_{\rm bar}$, confirm that the $m=2$ mode tumbles with substantially
different speeds.  On $\sim 1$\,pc scale,  $\Omega_{\rm bar}\sim 4\times 10^{-5}\,{\rm yr^{-1}}$, while on 
$\sim 0.01$\,pc scale, $\Omega_{\rm bar}\sim 5\times 10^{-3}\,{\rm yr^{-1}}$, about 2 orders of magnitude 
faster. 

While these pattern speeds are well-defined, there is no clear `material' separation between the spatial 
scales, i.e., no separation between bars. This continuity of the flow properties is associated with continuity 
of pattern speeds --- smaller scales
correspond to larger pattern speeds, $\Omega_{\rm bar}(r) \sim r^{-\alpha}$, with $\alpha \sim 1$. The 
parameter $\alpha$ is determined by the mass distribution $M(r)$ inside the collapsing gas at the onset of the 
central runaway.  The gas volume density scales as $\rho\sim r^{-2}$ in the central 
region (Fig.~\ref{fig:EllipDen1}) and dominates over the DM density distribution there, causing the gas to 
decouple dynamically from the DM background in the region of the central runaway. The pattern speed is given 
by $\Omega_{\rm bar}(r)\sim [M(r)/r^3]^{1/2}$, where $M(r)\sim r$, which explains the instantaneous value of 
$\alpha$ above.   

\begin{figure}
\centerline{
  \includegraphics[width=0.55\textwidth,angle=0] {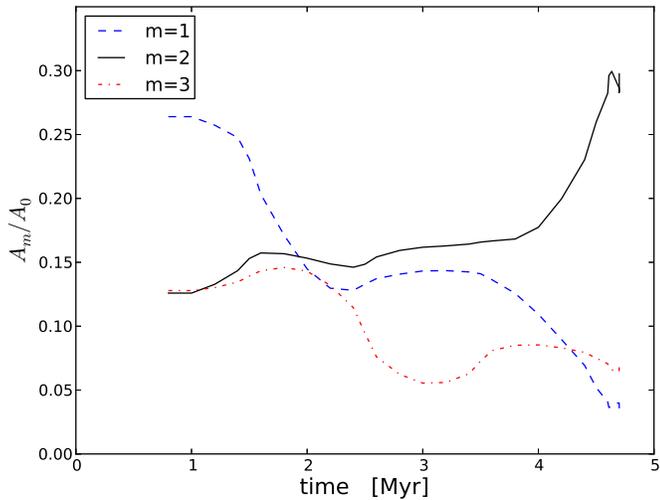}
}
\caption{Evolution of the Fourier amplitudes of modes with $m=1$, 2, and 3, normalized to the amplitude for 
$m=0$, for $r \leq 1$\,pc and $\Delta z = 0.25$\,pc about the equatorial plane.
}
\label{fig:Fourier}
\end{figure}

\begin{figure*}
\centerline{
  {\includegraphics[width=0.5\textwidth,angle=0] {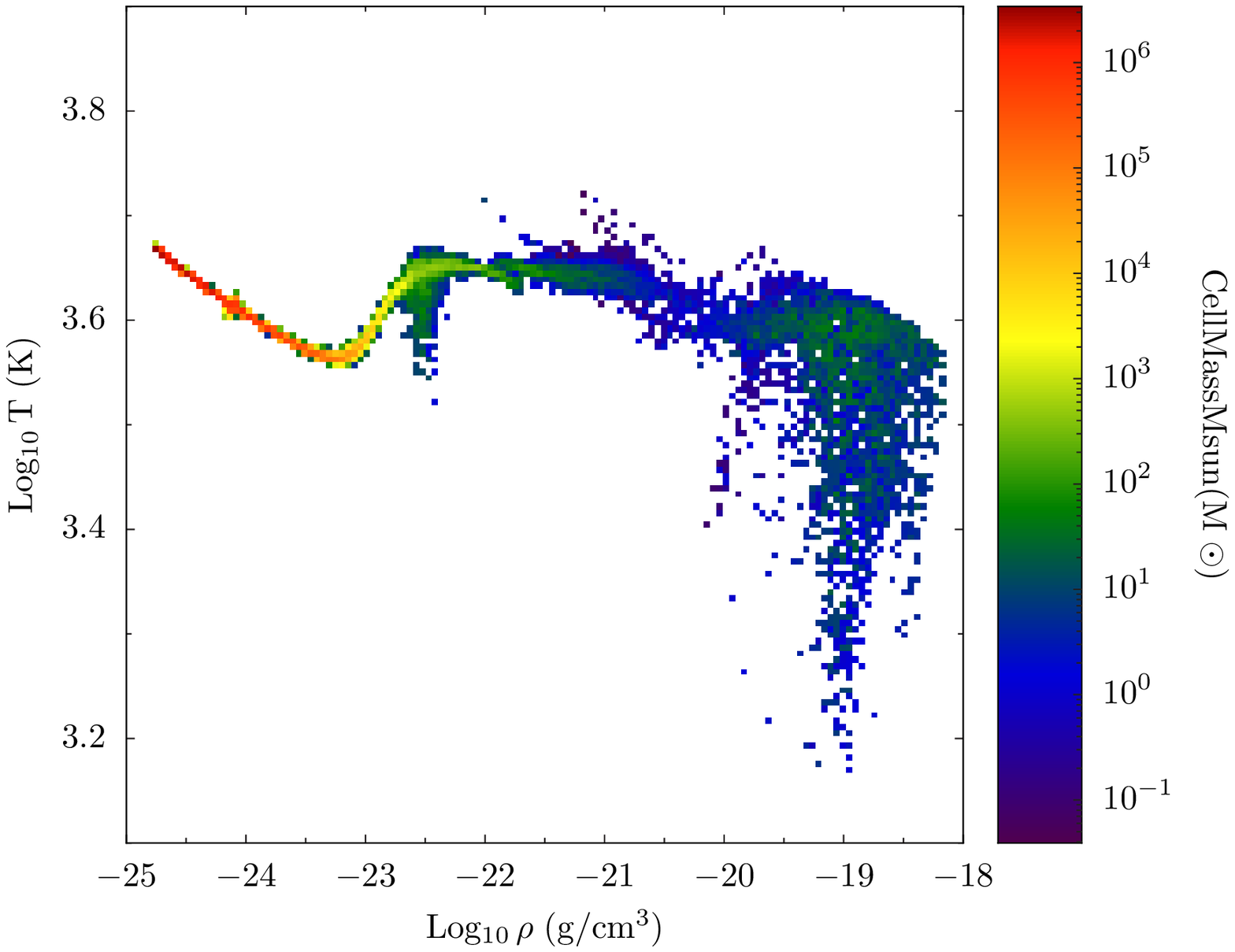}}
  {\includegraphics[width=0.5\textwidth,angle=0] {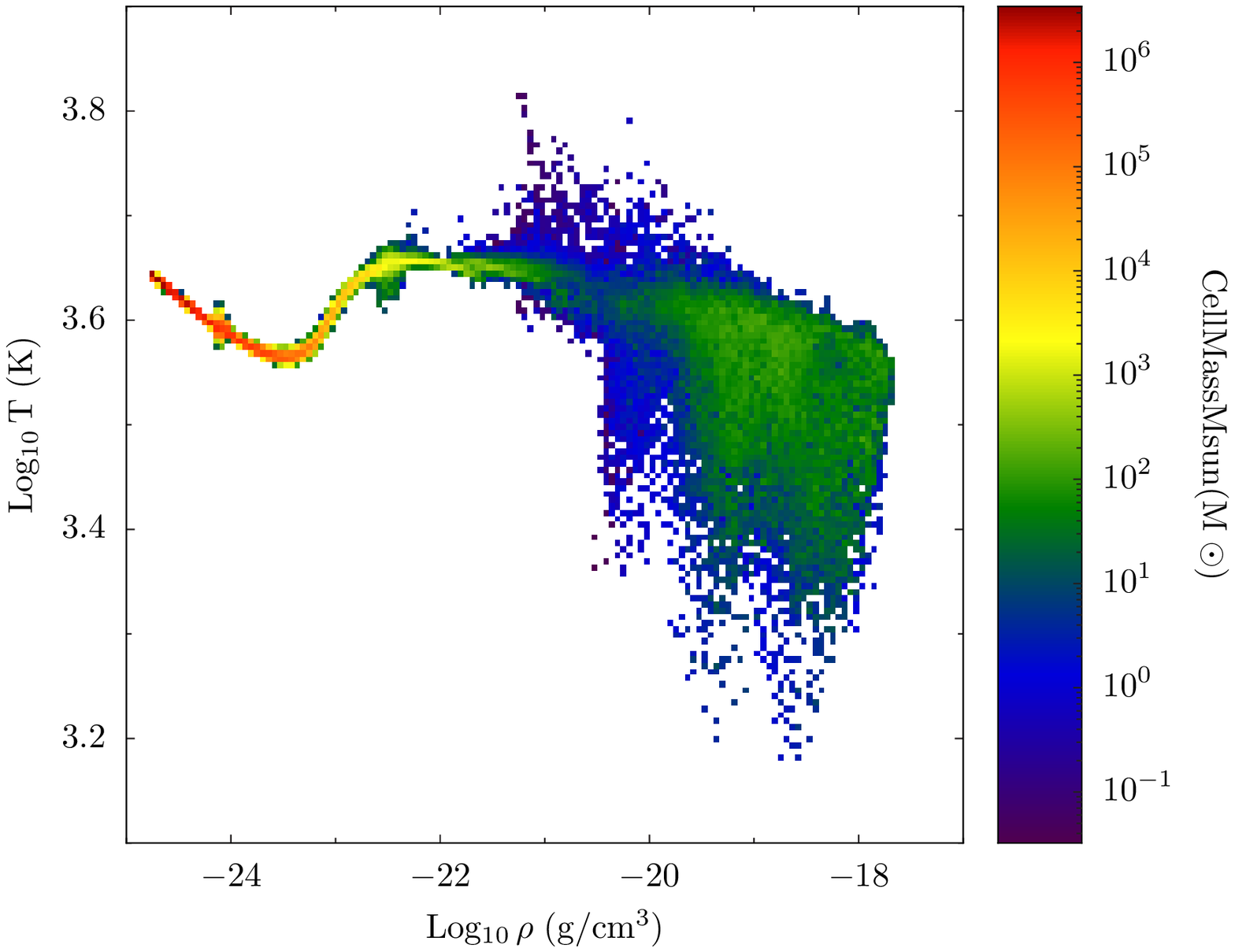}}
}
\centerline{
  {\includegraphics[width=0.5\textwidth,angle=0] {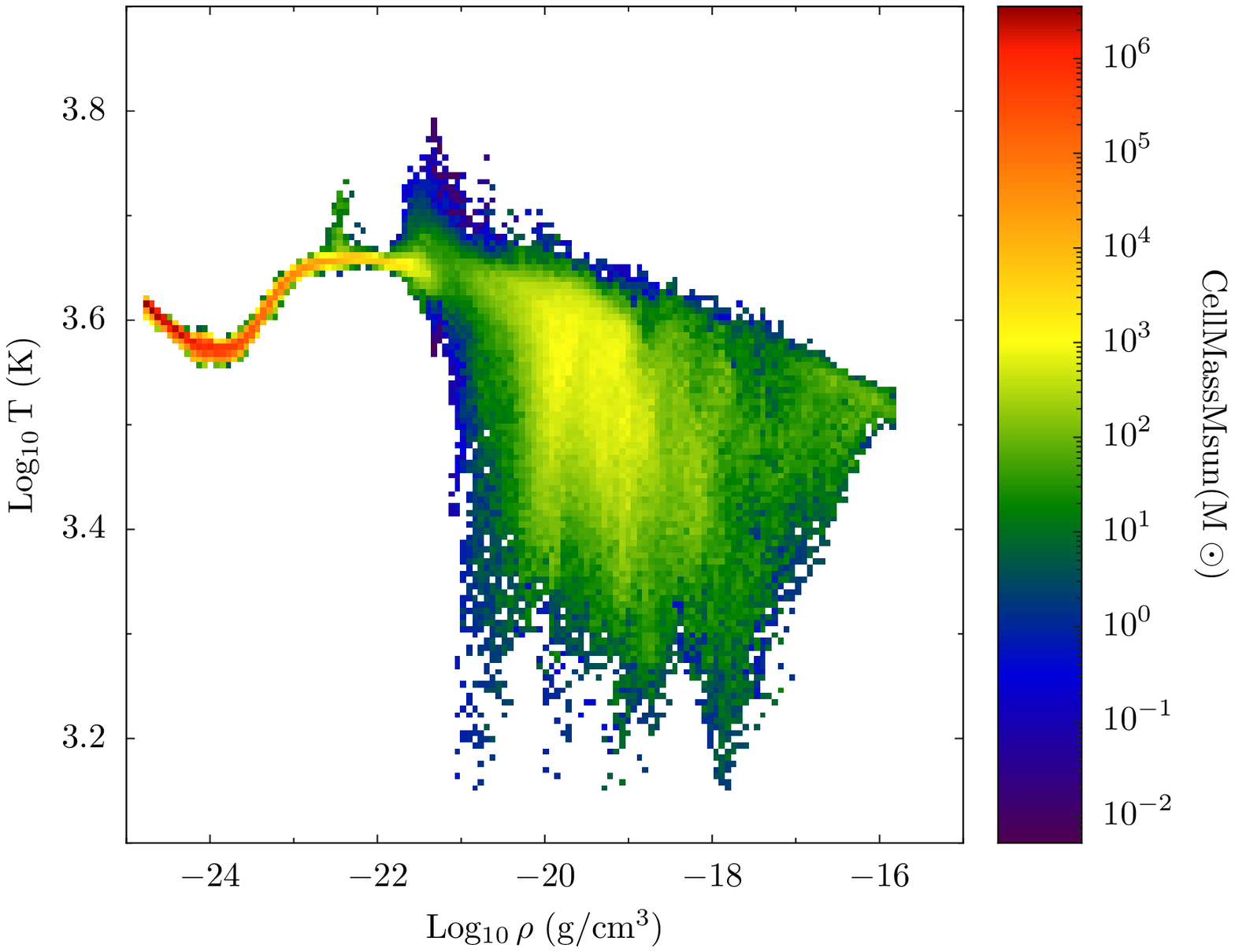}}
  {\includegraphics[width=0.5\textwidth,angle=0] {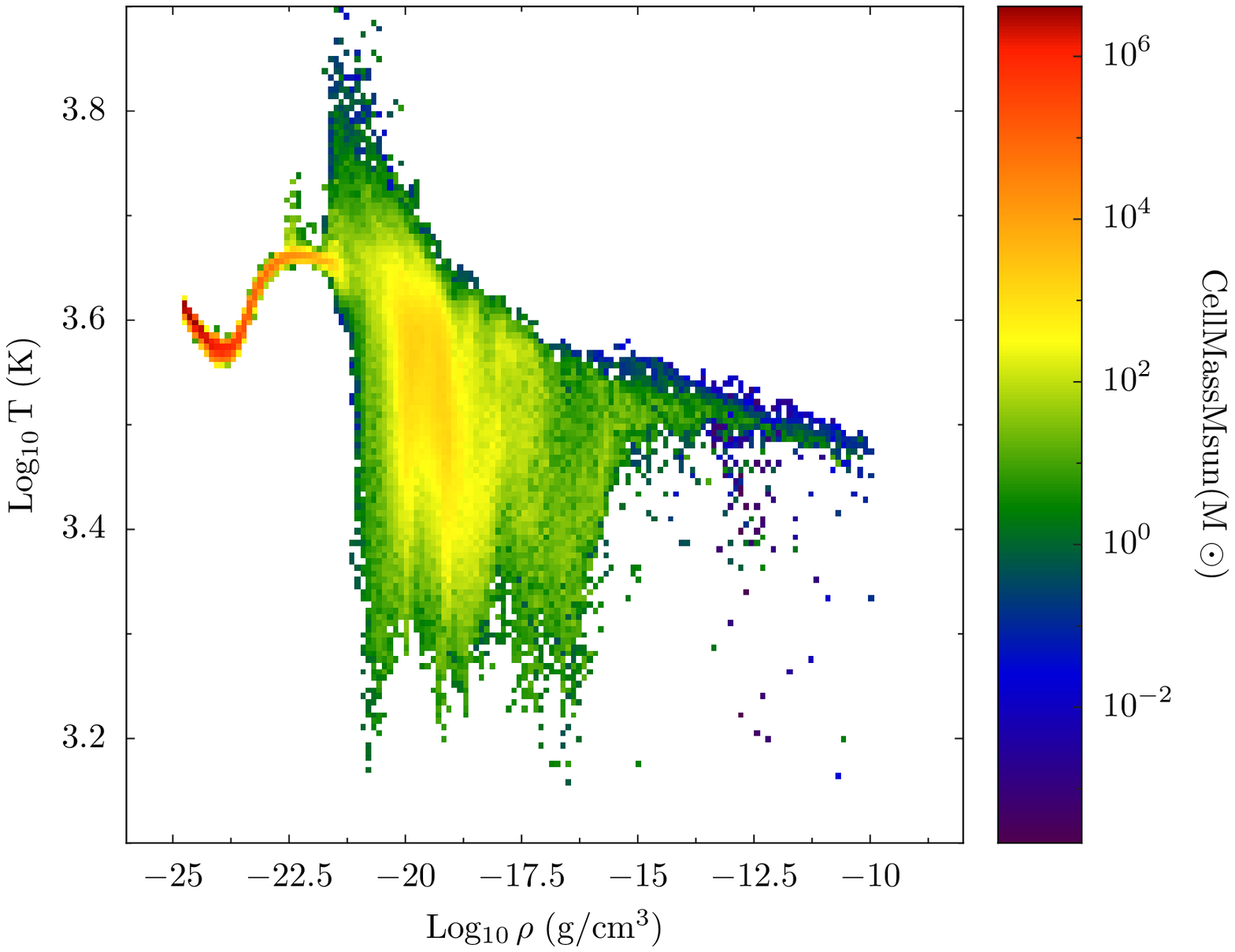}}
}
\caption{The temperature variation with the density in the collapsing flow of model B at four different times: 
{\it Upper left:} $t\sim 1$\,Myr, {\it Upper right:} $\sim 2$\,Myr, {\it Lower left:} $\sim 4.61$\,Myr, and 
{\it Lower right:} $\sim 4.70$\,Myr.  Only cells at 
spherical radius $R > 2\times 10^{-4}$\,pc are shown. The colors represent the frequency of cells in the 
respective mass range (right scale). Note the appearance of the radial shock which moves to lower densities
with time.
}
\label{fig:Temp_rho}
\end{figure*}

To verify that the bar-like ($m=2$) mode dominates over other modes, e.g., $m=1$ and 3, we have Fourier 
analyzed the gas response to the asymmetric potential.  The dominant mode at early times is the $m=1$ mode --- 
the disk center of mass is initially
perturbed by the asymmetry developed in the overall mass distribution (Fig.~\ref{fig:Fourier}).  This mode, 
however, decays quickly to a negligible amplitude.  An additional odd mode, $m=3$, is also present, although at 
lower amplitude --- it decays similarly. On the other hand, the even mode $m=2$ grows to a substantial 
amplitude with time.  At about $t\sim 4$\,Myr, it enters exponential growth --- this explains the appearance of 
the gaseous bar at this stage. Clearly, this bar-like mode has formed early in the evolution, when the gas 
component does not dominate the potential at any radius, and when the global stability parameter $\alpha < 0.34$ 
(see Section~\ref{sec:AngMom}). This mode is stimulated by the overall mass 
distribution. The exponential growth at later time happens exactly when and where the gravitational potential 
of the gas becomes the dominant one.  Thus the bar-like mode appears to be driven initially by the shape of the 
overall potential, but subsequently runs away due to the self-gravity of the gas.

\begin{figure*}
\centerline{
    \includegraphics[width=0.53\textwidth,angle=0] {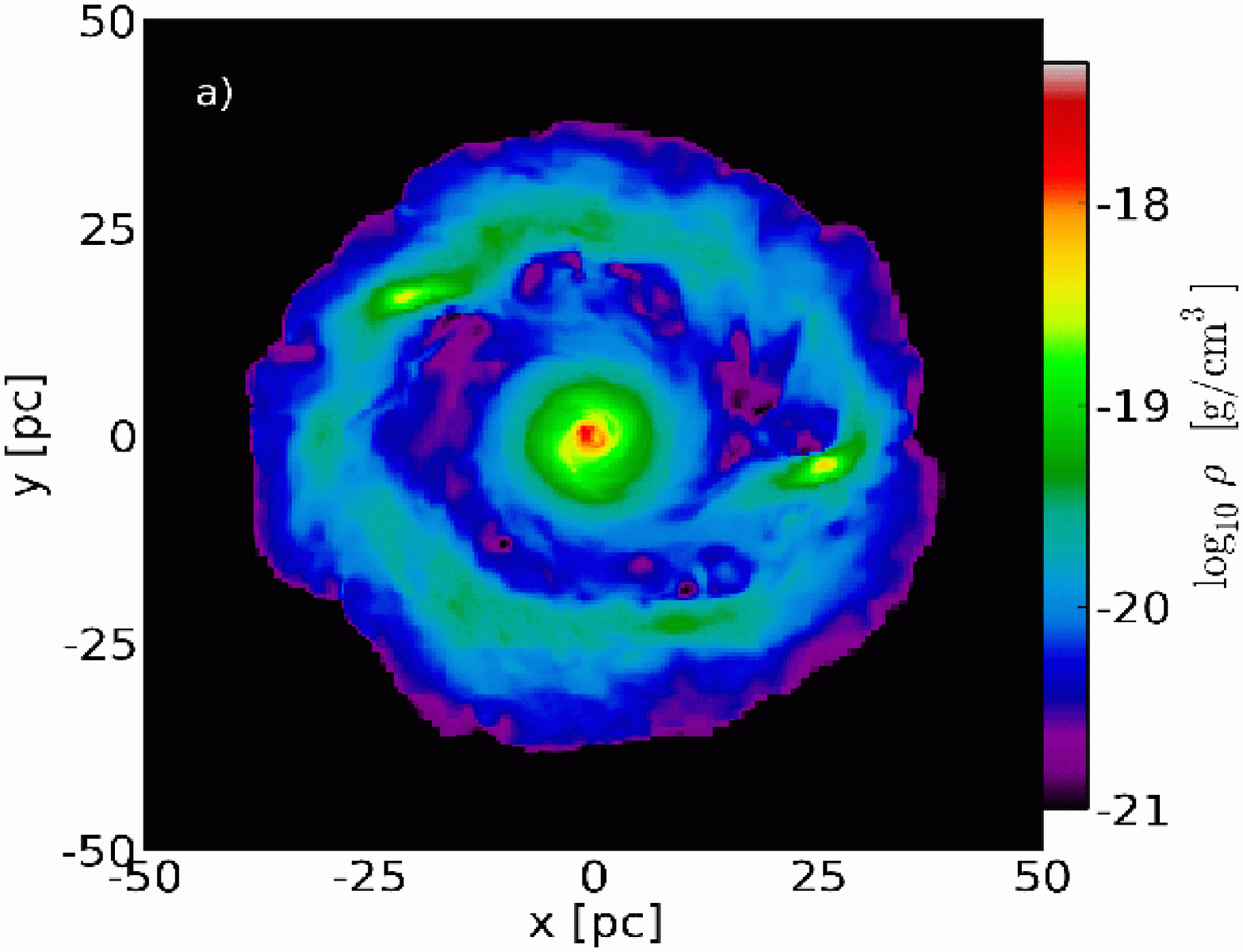}
     \includegraphics[width=0.53\textwidth,angle=0] {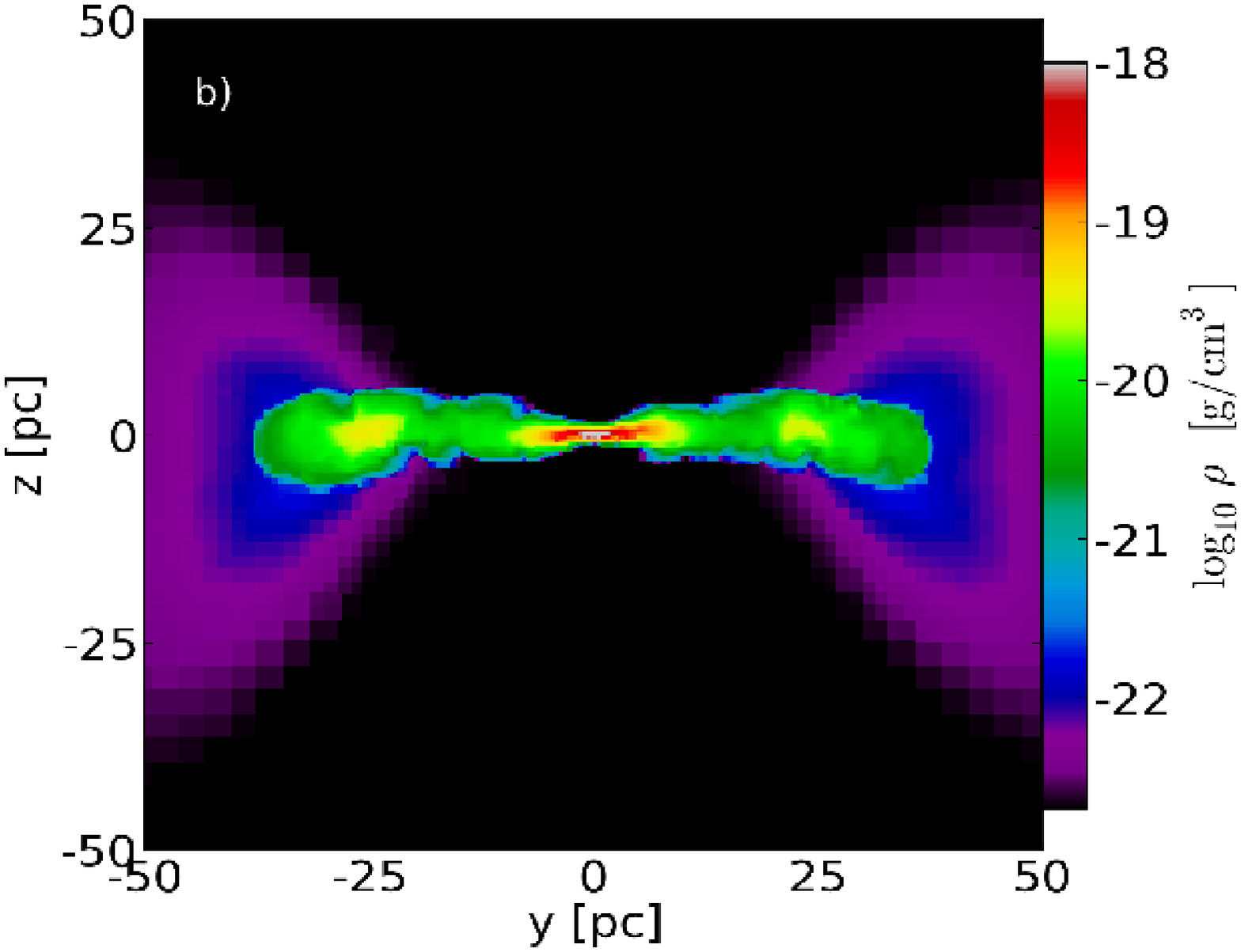}
}
\caption{
{\it (a)} Face-on density-weighted projection of density of the gas disk forming from collapsing gas at 
$t=13.2$\,Myr in Run D. Box size is 100\,pc and its vertical extent $\Delta z = 2$\,pc.
{\it (b)} Edge-on projection of of the same disk.  The disk edge is clearly visible, as well as the torus 
supported by turbulent pressure and delineated by shocks. The anisotropic density distribution is beginning to 
take shape and extends well beyond the torus.
}
\label{fig:frame13.2}
\end{figure*}

The temperature variations with density for model B are shown in Fig.~\ref{fig:Temp_rho}. At the start of disk 
formation, a radial shock forms at $r\sim 2$\,pc and then gradually propagates out to 10\,pc. At this time the 
central runaway is triggered. As shown by Fig.~\ref{fig:Temp_rho}, the posthock gas rapidly cools down to low 
$T$ in the postshock region because of the high density. Within the growing disk, $r\ltorder 10$\,pc, we 
observe an increasing range in $T\sim 1.5\times 10^3-10^4$\,K and $\rho\sim 10^{-17}-10^{-21}\,{\rm g\,cm^{-3}}$. 
The upper limit to the temperature is dropping with density. The central runaway at $\ltorder 1$\,pc narrows 
the $T$-range to mostly $T\sim 3\times 10^3$\,K but increases the range in $\rho\sim 10^{-15}-10^{-10}\,{\rm 
g\,cm^{-3}}$. The maximal postshock $T$ in the disk is slightly increasing with time, which reflects the 
increasing shock-impact velocity of the gas which comes from progressively larger distances.

\subsection{Interplay between off-center fragmentation and central runaway}
\label{sec:alternative}

As we noted in Section~\ref{sec:collapse},  
models A -- C exhibit a central runaway collapse, while models D and E do not show it --- not at 
$t=4.7$\,Myr nor at anytime later on. We now 
describe the evolution of the representative model D and analyze the reasons for its differences from model B.
We shall discuss additional models as well.

The first stage of collapse in model D proceeds similarly to model B, but leads to a disk within the 
centrifugal barrier that has a much lower surface density at the center.  Figure~\ref{fig:SurfDens} shows 
that this disk possesses a density core that grows to $\sim 2-3$\,pc at $t=4.7$\,Myr --- the time of the 
onset of the central runaway in model B, in sharp contrast with this model. In model E, the density core 
extends to $\sim 8$\,pc at this time. When the gas distributions in models B and D are compared before the 
central runaway in model B, at $t\sim 4.0$\,yr, the core density in B is higher by about 2 orders of 
magnitude than in D (Fig.~\ref{fig:profile}), and by even more compared to model E.

Clearly, a cuspier DM distribution leads to a more centrally concentrated gas distribution. This has a 
clear dynamical consequence: the dynamical timescale {\it prior} to the central runaway is shorter in more 
cuspy potentials. We observe that, at early times of the simulations, the $m=1$ mode is the dominant mode 
for all the models. This mode is damped faster in models with cuspier DM distributions, via dynamical
friction, presumably of the gas against the DM distributions. The $m=2$ mode, which is responsible for the
central runaway collapse, can only grow after $m=1$ has decayed significantly (Fig.~\ref{fig:Fourier}).
Therefore, our models A -- E form
a sequence along which the dynamical timescale becomes longer. Any dynamical instability will have a 
tendency to increase its amplitude more slowly along this sequence.  This includes the growth of $m=1$, 2 
and 3 modes that we observe, in different combinations, in these models.

\begin{figure}
\centerline{
  \includegraphics[width=0.53\textwidth,angle=0] {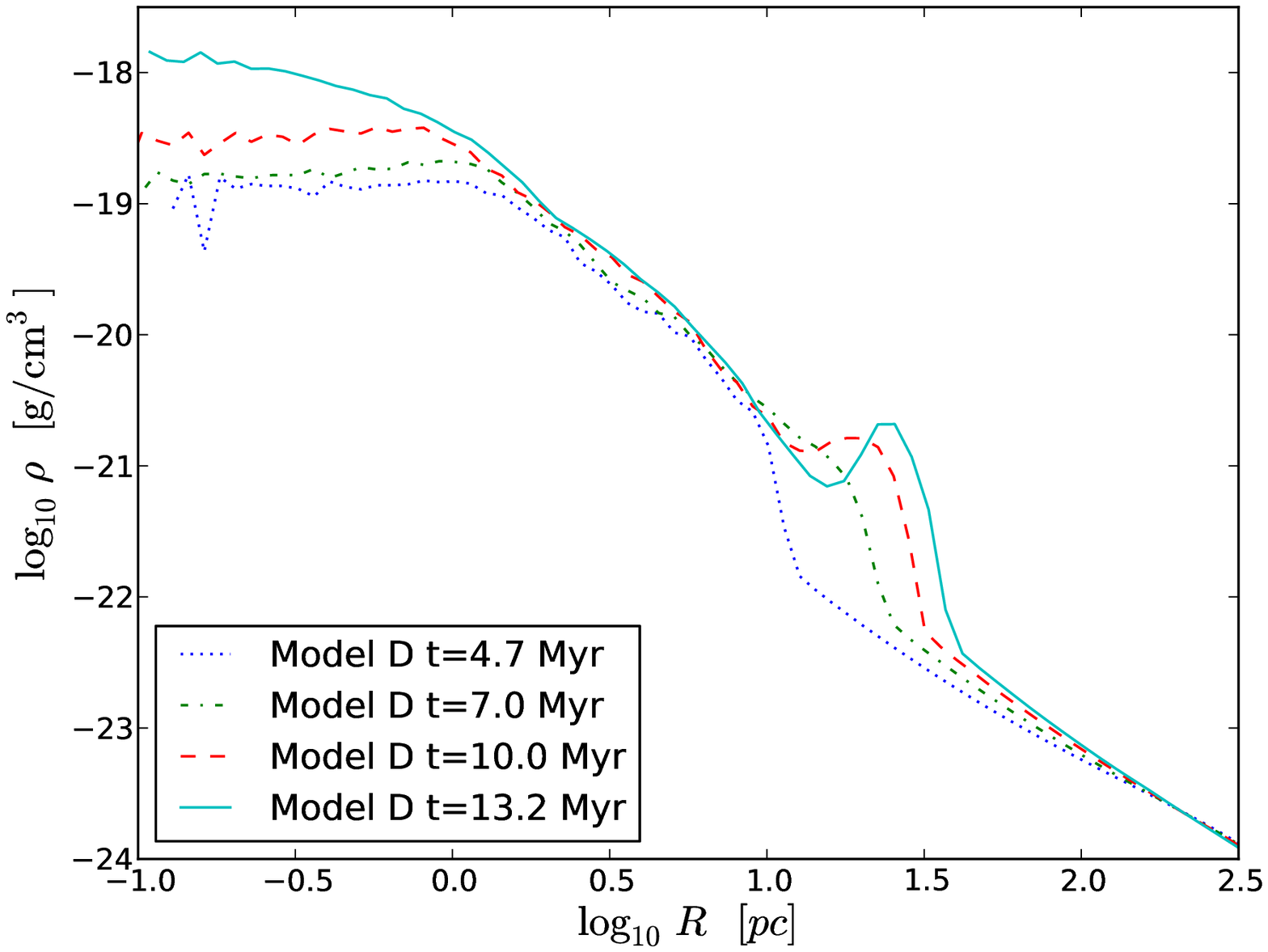}
}
\caption{
Evolution of the gas density profile, averaged over spherical shells, after $t=4.7$\,Myr, in model D. 
Comparing with Figures~\ref{fig:frame4.6} 
and~\ref{fig:profile}, the disk gets bigger and its central density increases.  Note the formation of the torus 
outside the disk, which shows mild fragmentation in Fig.~\ref{fig:frame13.2}.
}
\label{fig:DensRunD}
\end{figure}

\begin{figure*}
\centerline{
  {\includegraphics[width=0.5\textwidth,angle=0] {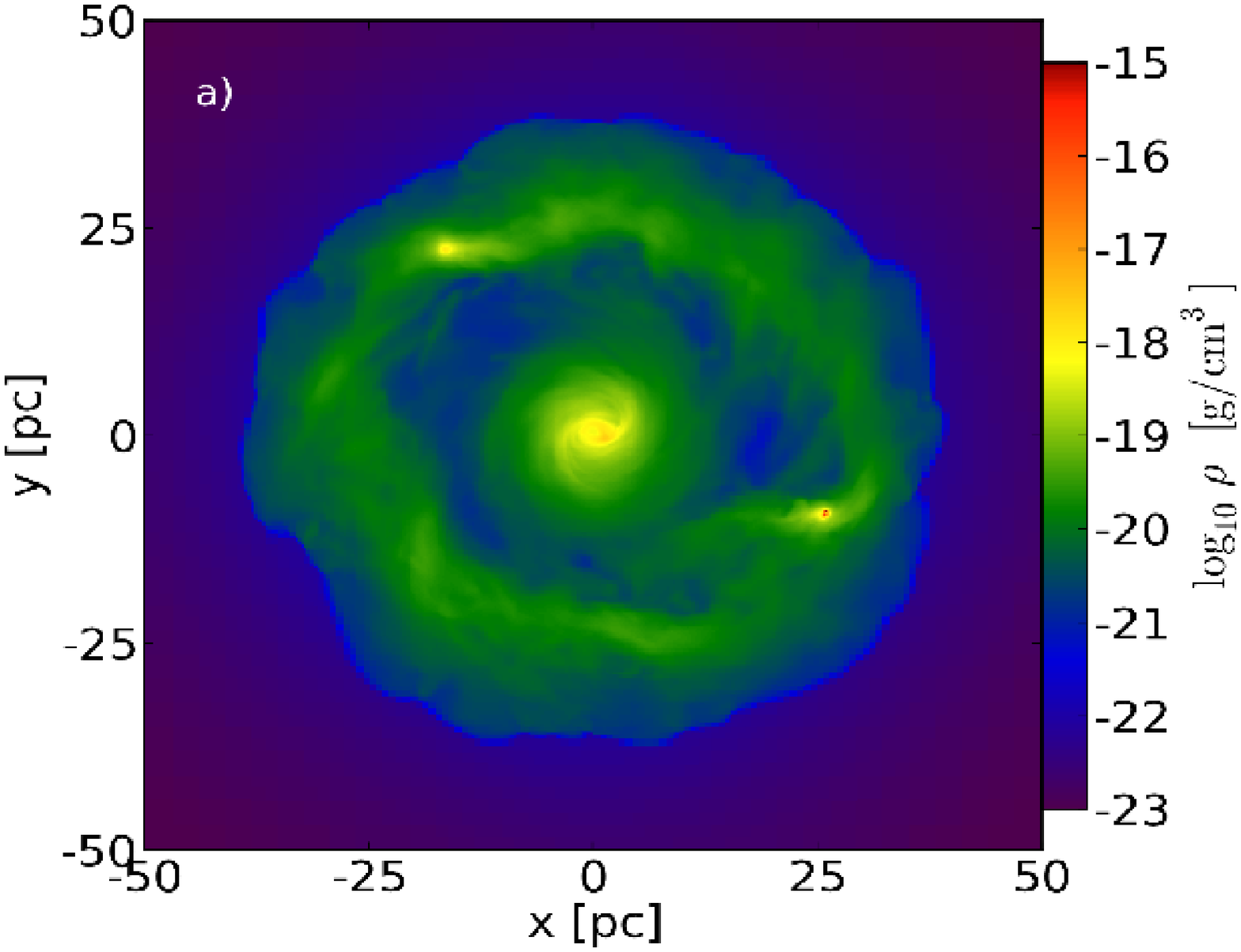}}
  {\includegraphics[width=0.5\textwidth,angle=0] {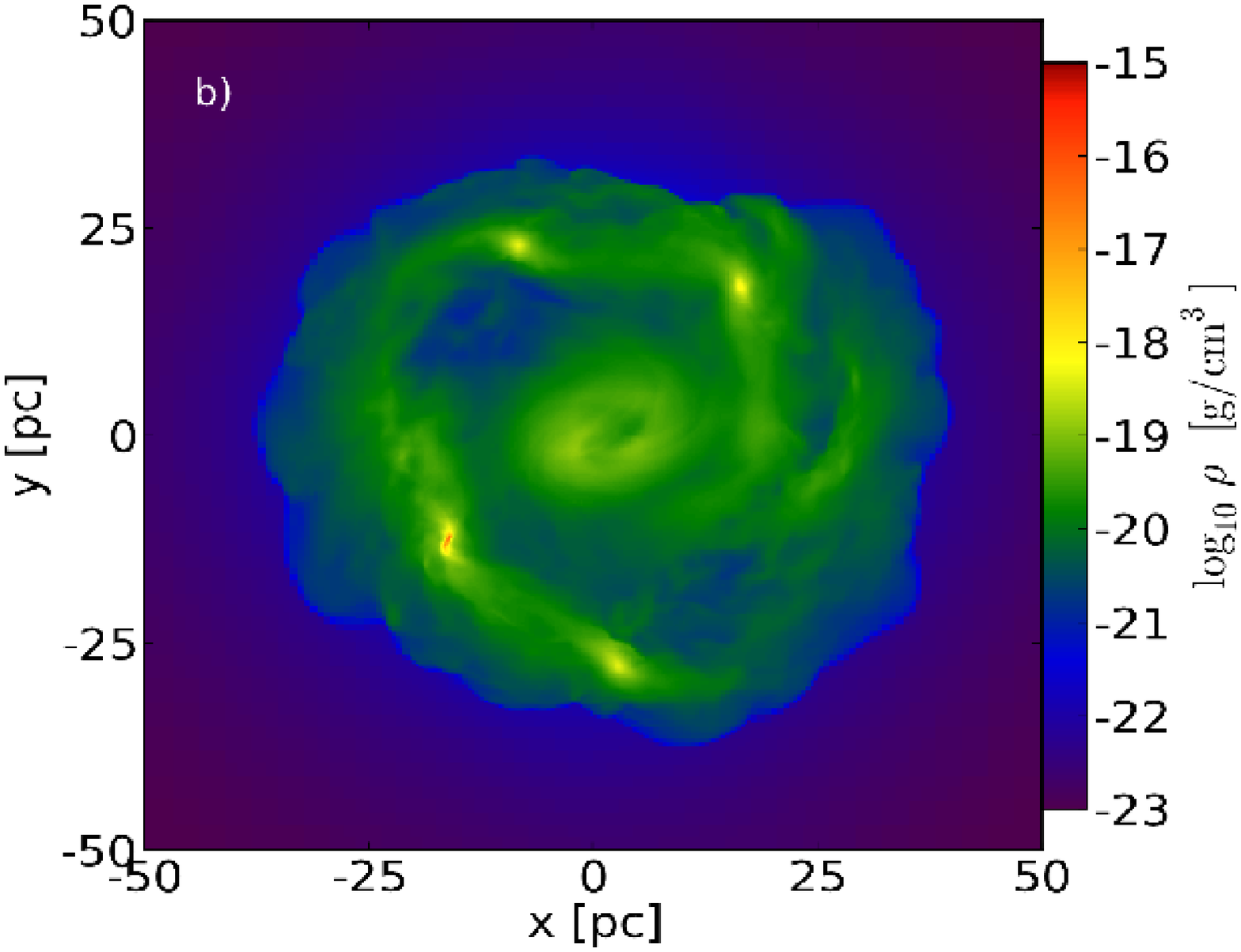}}
}
\caption{
Density-weighted projection of density of {\it (a)} model D, and {\it (b)} model E, in the equatorial 
plane with a thickness $\Delta z = 2$\,pc,
show the gas disk forming from collapsing gas at $t=13.4$\,Myr, when the tori in models D and E exhibit 
fragmentation.  Box size is 100\,pc.
}
\label{fig:TorusFrag}
\end{figure*}

\begin{figure*}
\centerline{
  \includegraphics[width=2.0\columnwidth,angle=0] {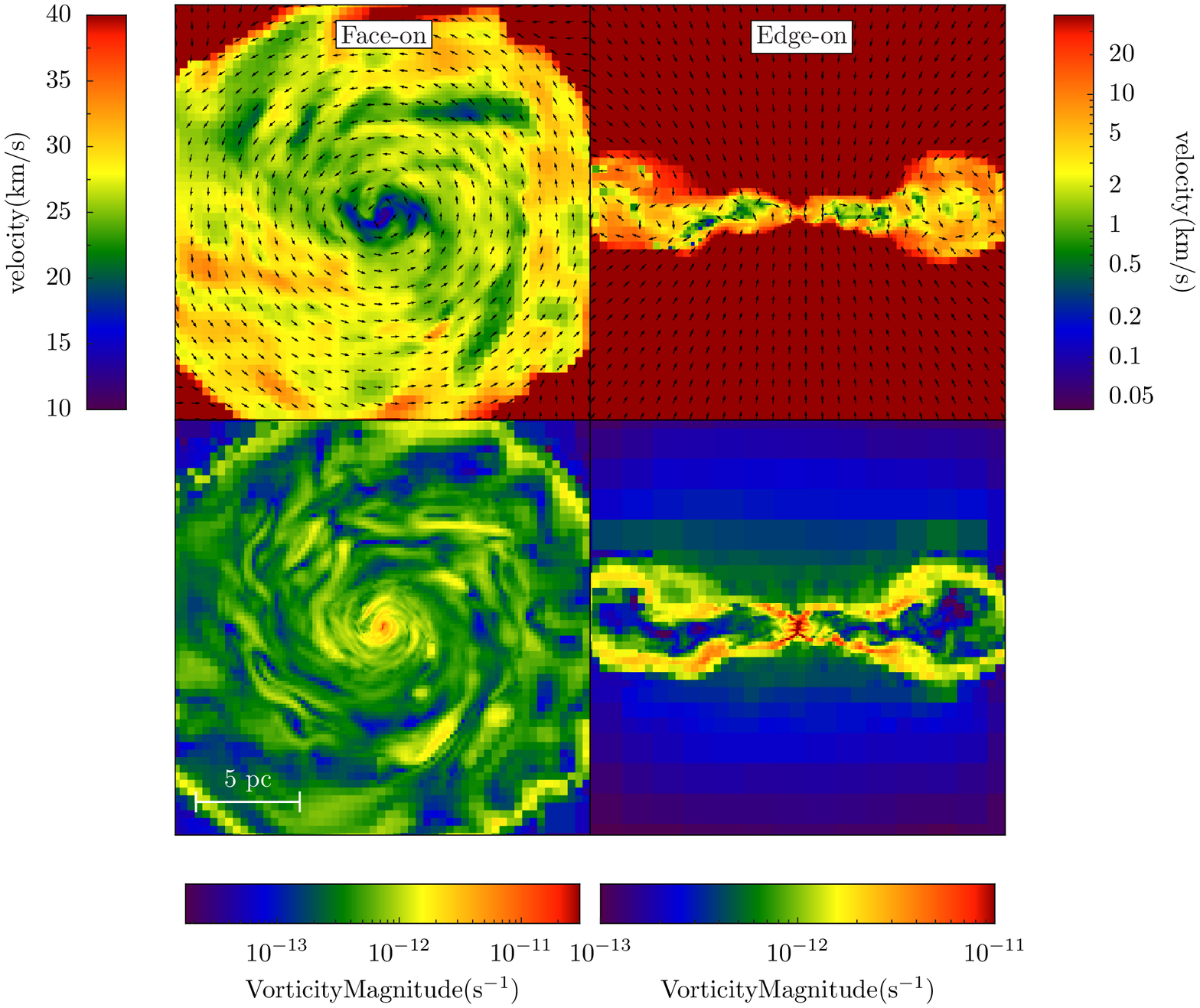}
}
\caption { {\it Top:} Face-on (left) and edge-on (right) projections of the velocity field in the collapsing 
flow of model B, on scales of 20\,pc. Arrows show the direction of the flow {\it only} --- the velocity values 
are given by 
the color palette.  While the face-on flow is dominated by the rotational component (at larger radii), the 
edge-on slice is dominated by turbulence. Positions of radial and surface shocks are clearly delineated by 
sharply increased turbulence. {\it Bottom:} Face-on (left) and edge-on (right) slices of the vorticity 
field, ${\rm\bf w}$, in the collapsing flow of model B on 20\,pc scales.  Vorticity is generated by the 
oblique shocks that delineate the disk, by spiral shocks within the disk, and by the central runaway.
}
\label{fig:outerTurb}
\end{figure*}

Further evolution of model D shows the growth of the disk behind the radial shock.  The central density 
increases by about an order of magnitude, but still falls short of the beginning of the central runaway in 
model B (Figure~\ref{fig:profile}). 
This disk growth is related to gas with an increasing angular momentum $j$, and steadily growing free-fall velocity, 
arriving from larger $r$.  After $\sim 6-7$\,Myr, the turbulent pressure in the shocked gas
substantially exceeds the thermal pressure behind the shock. This leads to the formation of a 
geometrically-thick torus with a growing surface (and volume) density (Fig.~\ref{fig:frame13.2}). 
Figure~\ref{fig:DensRunD} displays the formation and evolution of this torus at $r\sim 20-100$\,pc. The 
density profile at $t=13.2$\,Myr shows that a significant fraction of the collapsing gas has stagnated in 
the torus. At this time, the gas in this region is mainly in circular motion --- the radial motion 
essentially ceases and the turbulent motions decay. The torus becomes azimuthally inhomogeneous, and 
increasing density leads initially to mild off-center fragmentation, but most of the fragments are 
immediately sheared and destroyed (e.g., Fig.~\ref{fig:frame13.2}a). Finally, at $t\sim 13.4$\,Myr, one of 
the fragments becomes substantially compact and begins gravitational collapse and the simulation is 
stopped (Fig.~\ref{fig:TorusFrag}). Model 
E is similar to D, with off-center fragmentation in the torus occurring at the same time, $t\sim 13.4$\,Myr.

We have followed the evolution of these fragments in models D and E. It differs substantially from the
evolution of the central runaway (i.e., collapse of the central fragment) in models A -- C. The typical
mass of the fragments is about $10^4\,M_\odot$, but it is not clear if this mass is characteristic of 
the off-center runaway. Each of
the fragments in D and E collapses and exhibits a fission into two fragments. We did not follow their evolution
further.

The evolution of model D thus diverges from that of model B.  The appearance of fragments perturbs the central 
region of the gas disk.
We observe that these perturbations lead to a loss of symmetry in the disk as well as displacing its center of 
mass, depending on the number of fragments and their azimuthal distribution.  The formation of an even-mode 
bar is suppressed.  Therefore, the effect of fragmentation in the torus is that the bar instability is damped.

This result reveals competition between two timescales --- that of the central runaway and that of off-center 
fragmentation in the models.  The development of the bar cascade and of off-center fragmentation are 
facilitated by different instabilities. Gas inflow, of course, drives both instabilities, but the DM plays an 
important role by regulating the timescale of the central runaway, as well as the decay timescale for the $m=1$
mode.  Gaseous bar formation is a global 
instability in the gas disk and appears to be triggered by both the gas gravity and the global distribution of 
the collapsing matter. In more centrally-concentrated disks, this instability happens earlier. On the other 
hand, the 
fragmentation in the torus is determined by the local surface density and the local pressure --- this is a 
local instability. If the gas turbulent velocities decay first in the torus (i.e., in the outer disk), the
torus fragments first. The competition between
these two instabilities, local and global, has been investigated in \citet{Begelman.Shlosman:09}.

To summarize, we find that the DM density profile, specifically its cuspiness, plays the role of the 
discriminator between models that experience the second stage of gravitational collapse past the centrifugal 
barrier and models that do not exhibit this dynamical stage.  This dichotomy is related directly to the surface 
density profile of the growing disk within the centrifugal barrier. But, as we have shown in 
Section~\ref{sec:collapse}, it also depends on the asymmetry of the outer mass distribution that develops in 
the collapsing matter, triggering the bar cascade.  Essentially, this corresponds to the dynamical decoupling 
of the
gas from the underlying DM potential.  During the initial stage of gas collapse, the DM potential dominates at 
all radii and the baryon density is lower than the DM density everywhere. However, since the DM potential of 
model B is more cuspy than that of model D (Table~\ref{tab:table1} and Fig.~\ref{fig:SurfDens}), its gas 
contracts more, and the surface density of the forming disk is substantially higher.  The disk continues to 
grow as a result of ongoing accretion and the settling down of the turbulent gas. We, therefore, turn to the 
details of this turbulent dynamics.

\subsection{The role of turbulence}
\label{sec:turbulence}

ENZO has been extensively tested to handle supersonic turbulent motions in a uniform density background
\citep[e.g.,][]{Kritsuk.etal:07,Kitsi.etal:09,Padoan.etal:09,Kritsuk.etal:11comp} and in stratified 
densities \citep[][]{Kritsuk.etal:11}.

Figure~\ref{fig:VrMach} shows the radial and tangential velocity profiles of the flow in Mach numbers, 
calculated using the velocity and temperature maps for model B. The right-hand maximum corresponds to the 
initial stage of the gravitational collapse and exceeds $\Mach\sim 3-5$.  The left-hand maximum reflects the 
accelerated runaway in the central region and exhibits the same range of Mach numbers. Both the radial and 
tangential flows are clearly supersonic. 

The upper frames of Figure~\ref{fig:outerTurb} show the velocity field for model B within the central 10\,pc 
at the time of the central runaway. The face-on disk displays spiral shocks in the gas, while the velocity 
field is dominated by rotation. At the same time, the edge-on disk shows a turbulent velocity field and a 
number of eddies. The geometrically-thin disk appears substantially puffed-up behind the radial shock. 
Because the temperature maps reveal that the shocks are nearly isothermal, the concurrent increase in the 
vertical thickness of the disk can come only from the turbulent flow behind the standing radial and 
vertical shocks.  Indeed, our estimates of thermal pressure gradients in the $z$-direction reveal that 
thermal pressure gradients are insufficient to puff up the disk to the extent seen in the upper right 
frame of Figure~\ref{fig:outerTurb}. 

While turbulence is notoriously difficult to define, we follow the definition which relies on the 
vorticity ${\rm\bf w} = {\rm\bf \nabla\times v}$, and its cross product with the velocity field, 
i.e., the inertial vortex force ${\rm\bf v \times \nabla \times v}$.
To quantify the turbulence within the disk, we have followed the vorticity field within the
computational box (lower frames of Fig.~\ref{fig:outerTurb}). As expected, on larger spatial scales 
the velocity field is irrotational
due to the relaxed initial conditions used in our simulations. As the inflow velocity grows with 
time and the velocity field becomes less regular, the vorticity increases and exhibits a 
discontinuity at the standing shock which envelops the disk. The postshock flow shows a sharp 
increase in the vorticity, which decays toward the equatorial plane.  
In the postshock region, the turbulence is transonic (Table~\ref{tab:table2}). It provides support for the 
vertical disk structure. Its decay, when moving radially-inward from the shock, results in the sharp decrease 
of the disk vertical thickness at smaller radii.

The vorticity in the disk 
appears to be driven by the spiral shocks and by the bar-like perturbation at the center (at later 
times) --- the spiral arms around this perturbation are turbulent as well, as can be seen in the 
face-on disk region.  The central region, which experiences the runaway collapse, exhibits the 
highest vorticity.  

\begin{figure}
\centerline{
  \includegraphics[width=0.55\textwidth,angle=0] {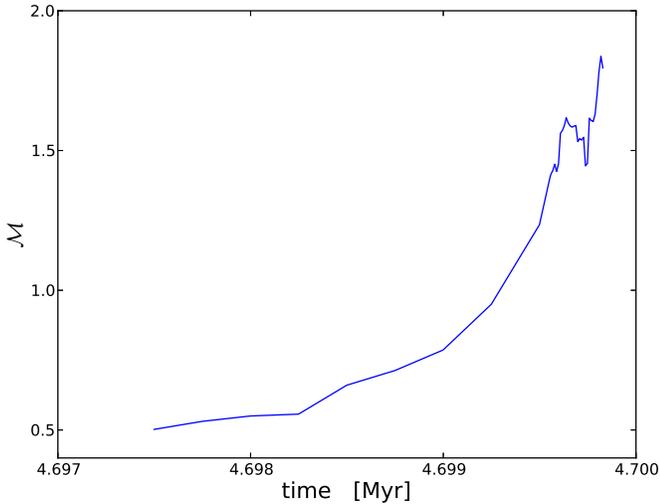}
}
\caption{Evolution of turbulent velocities in the central runaway region, in units of Mach number, in 
model B. The region sampled consists of a sphere positioned in the equatorial plane at 
$r_{\rm c} = 1,500$\,AU from the center, with a radius of 1,450\,AU. This corresponds to the first line 
in Table~\ref{tab:table2}.
} 
\label{fig:TurbCtr}
\end{figure}

We have also sampled the turbulent velocity field (in terms of the Mach number) on smaller spatial scales, 
at the time of the central runaway, using spherical sampling volumes whose positions in the disk plane at 
radii $r_{\rm c}$ are given in Table~\ref{tab:table2}. 
The evolution of the turbulent velocity within the central sphere (i.e., first line in 
Table~\ref{tab:table2}) is also displayed in Figure~\ref{fig:TurbCtr}.

\begin{table}
\caption{\bf Sampling the Turbulent Velocities in model B}.
\centering
\begin{tabular}{lcccccccc}
\hline
Center at $r_{\rm c}$ (pc) & Sphere Radius (pc)& $\Mach$ \\
\hline
 $7.2\times 10^{-3}$    &  $7.0\times 10^{-3}$  & 1.85 \\
 $7.5\times 10^{-3}$    &  $7.2\times 10^{-3}$  & 1.62 \\
 $1.0\times 10^{-1}$    &  $5.0\times 10^{-2}$  & 0.72 \\
 1.0                    &  0.3                  & 0.48 \\
 2.5                    &  1.0                  & 0.63 \\
 5.0                    &  2.0                  & 0.94 \\
\hline
\end{tabular}
\tablecomments{$r_{\rm c}$ is the distance from the rotation axis of the center of the sampling sphere (
see text for additional explanations). Last column: the estimated Mach number within the sampling sphere.
}
\label{tab:table2}
\end{table}

\begin{figure}
\centerline{
  \includegraphics[width=0.55\textwidth,angle=0] {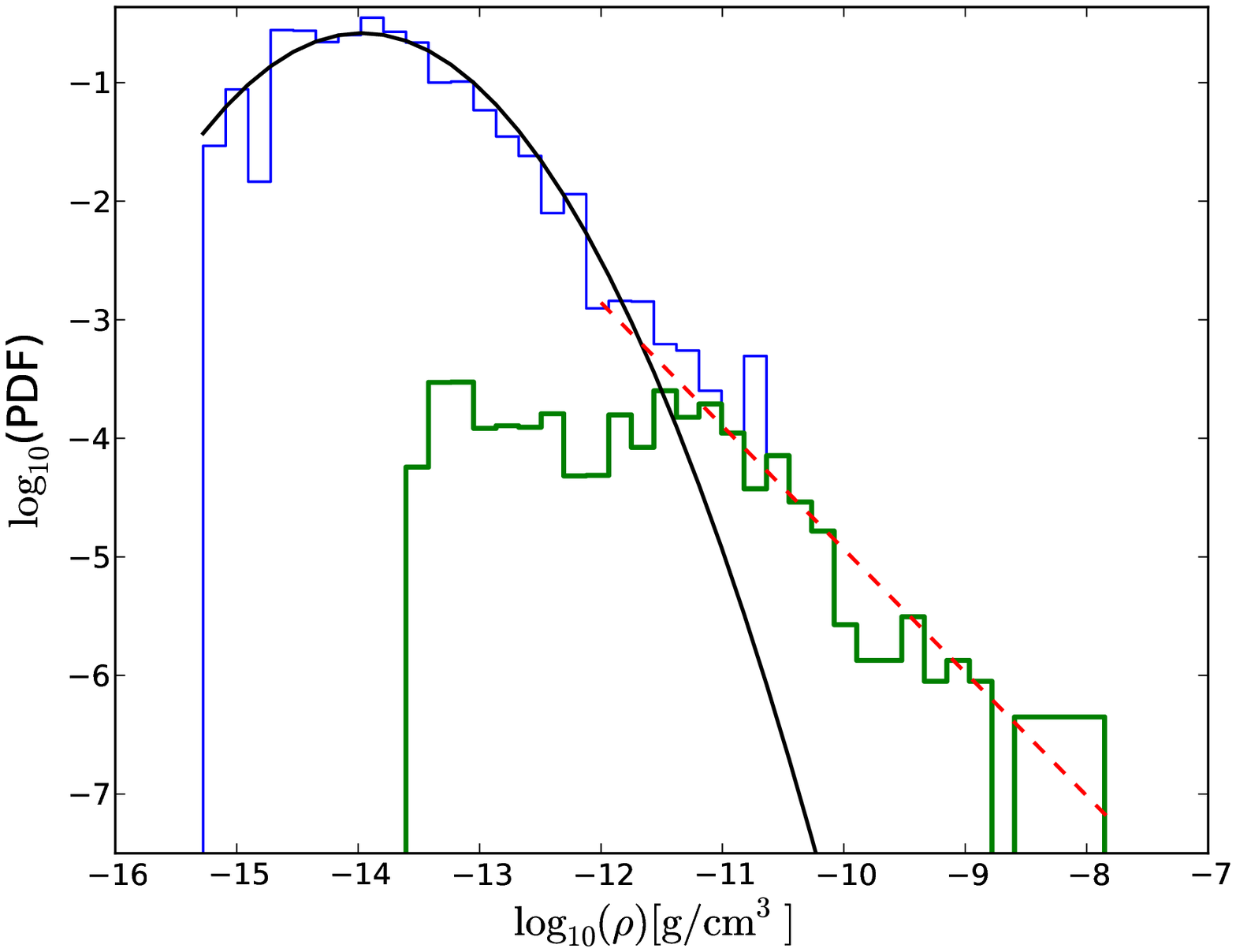}
}
\caption{The volume-averaged PDF of the gas density as a function of log$_{10}\,\rho$ measured during 
the central runaway at $t=4.7$\,Myr for model B and sampled with $\sim 1,500$ AMR cells.
The sampling shows the PDF of the central shell of radius 20\,AU--200\,AU (blue histogram) and the 
central sphere of 20\,AU (green histogram). Shown also are the lognormal 
fit (black) with dispersion $\sigma\sim 1.52\Mach$ for the blue histogram using Equation~\ref{eq:pdf} and 
the power law tail (red) for the green histogram with a slope of $\sim -1$. The collapsing gas has been 
sampled at the resolution of 1\,AU and the density fluctuations extend over 7.5 
decades.  The average baryon density in the sample sphere and shell is 
$<\rho>\sim 4.7\times 10^{-14}\,{\rm g\,cm^{-3}}$. 
} 
\label{fig:pdf}
\end{figure}

An alternative method for measuring the properties of supersonic turbulence is the PDF of the gas density. 
As discussed in Section~\ref{sec:theory}, the density PDF for homogeneous supersonic turbulence is
expected to follow a lognormal distribution 
\citep[e.g.,][]{Vazquez-Semadeni:94,Padoan:95,Padoan.Nordlund:02,Krumholz.McKee:05,Federrath.etal:11}.  To 
estimate the PDF 
in different regions of model B, we measure the grid densities within sampling spheres centered at various 
points of interest.  The sampling of the central region is centered at (0, 0, 0).  For each sampling center, 
we choose the radius of a sphere to enclose $\sim$\,1,500 AMR cells. 
These spheres sample different dynamical states of the collapsing gas: 
the central runaway, the outskirts of the disk, the transition region between the disk and the halo, and 
the gas at large. We carefully choose the size of these spheres to sample a large enough number of grid 
cells to obtain reasonable statistics, without mixing different dynamical regions in a single sphere. 

Figure~\ref{fig:pdf} shows the volume-averaged density PDF in the central region of model B during the 
central runaway, at $t=4.7$\,Myr. 
We are unable to fit a single analytic lognormal PDF, for a given mean density 
(Equation~\ref{eq:pdf}), to the entire density range. Instead, we use a combination of a lognormal and a
power-law distribution, with the power-law tail dominating at higher densities. At lower densities, 
comparison between the measured density PDF and the analytical lognormal PDF shows excellent agreement,
with the lognormal PDF fit extending over 4 decades in $\rho$. For  ${\rm log}\,\rho > -12$, the best fit 
is a power law tail with the slope of $\sim -1$ for about 3.5 decades in  ${\rm log}\,\rho$.  No such tail
is detected for other locations of sampling spheres. 

Such a power-law tail has been observed previously in two-dimensional simulations of homogeneous,
supersonic hydrodynamical turbulence in its early stages, before the formation of self-gravitating clumps, 
i.e., before fragmentation \citep{Scalo.etal:98}.  
This stage is similar to the present models, albeit with an important difference --- our central region is 
also in a state of a supersonic gravitational collapse and therefore has developed a substantial density 
gradient. In 
comparison, the \citet{Scalo.etal:98} power-law tail has a measured slope of $\sim -2$ before the 
self-gravitating clumps have formed, and a slope of $-1$ when these clumps are present.  In this respect, 
the central runaway in our models agrees well with the similar dynamic state in the gas of Scalo et al. --- 
it corresponds to the gravitational collapse of a ``fragment."  Such a power law is predicted for the 
solutions of Burgers equation (which is pressureless) at high densities \citep[e.g.,][]{Gotoh.Kraich:93}.
Power-law tails have also been observed in three-dimensional simulations \citep[e.g.,][]{Federra.etal:08,
Kritsuk.etal:11}.

In a more recent work, \citet{Kritsuk.etal:11} has targeted isothermal supersonic turbulent flows in the presence
of gas self-gravity, for the purpose of determining the mass density PDF. A random force has been used to drive
the turbulence on large spatial scales, in contrast with our models where turbulence is driven by gravitational
collapse only.  Despite these differences, we are in agreement that the power-law tail in the PDF appears at the
time of the central runaway, when and where the {\it local} gravity in the gas becomes important, albeit the
slopes are different at the end of the simulations: $-1.7$ for \citet{Kritsuk.etal:11} and $-1$ for our models
A -- C. Sampling away from the central runaway site in our models does not show the power-law tail, indicating
that there are no other self-gravitating fragments within the computational box.
 
Why is the power-law slope shallower in our simulations, i.e., $-1.7$ vs $-1$?  \citet{Kritsuk.etal:11} comment
that a shallower, $-1$ slope does appear at high densities and associate this with the mass pile-up resulting
from dynamically important angular momentum in the region. Recall that the central runaway in our simulations is 
angular momentum-dominated, as we show in Section~\ref{sec:collapse}. 

We have tested the origin of this power-law PDF by sampling the region with concentric spheres having
progressively smaller radii. We find that the range fit by the lognormal PDF shrinks along this sequence,
primarily because of cutting off the lower densities, while 
the high-density end remains untouched (compare the green and blue histograms in Figure~\ref{fig:pdf}). 
The high-density 
end of the power-law PDF, therefore, corresponds to the very high central density in our simulations, and 
the density stratification in the central region is responsible for the $-1$ slope power-law PDF. 
Clearly, understanding of the power tail in the PDF requires additional theoretical analysis.

In our models discussed here, the initial conditions are simplified and do not include 
turbulent motions. 
This delays the onset of turbulence, which takes time to fully develop in the outer part of the flow. In more 
realistic models, the infalling gas is expected to be already partly turbulent. However, one can also argue 
in favor of initially laminar and mildly sub-Alfvenic flow in minihalos
\citep[e.g.,][]{Abel.etal:02,Bromm.Larson:04,Yoshida.etal:06,Greif.etal:09,Schleicher.etal:09}. As the infall 
develops, 
turbulence sets in spontaneously and greatly increases after the gas has crossed the standing shock enveloping 
the disk within the central few parsecs.  Turbulence which has developed during gravitational collapse has 
been noticed by other authors \citep{Levine.etal:08,Wise.etal:08,Regan.Haehnelt:09}, and its impact on the 
gravitational stability of the flow has been analyzed by \citet{Begelman.Shlosman:09}.  As discussed in 
Section~\ref{sec:theory}, the effect of turbulence developing during gravitational collapse is to suppress 
gas fragmentation.

The absence of fragmentation in our simulations of collapsing flow is evident in Figure~\ref{fig:frame4.6},
which provides a snapshot of the face-on disk in model B. A simple estimate based on the floor temperature of
the gas with a primordial abundance, e.g., $T\sim 10^4$\,K, within a $2\times 10^8\,M_\odot$ DM halo shows 
that the ratio of the gas mass within some spherical radius $R$ within this halo to the Jeans mass 
at this temperature and density is of order $\sim 3-4$.  Because the Jeans mass depends on the rms velocity 
to the third power, this ratio will decrease to $\sim 1$ in a flow with transonic turbulent velocities.

\subsection{Randomizing gas motions: model Dmod}
\label{sec:modelDmod}

We have rerun model D with less organized baryonic initial momenta (Section~\ref{sec:alternative} and 
Table~\ref{tab:table1}).
The purpose of this run is to introduce more realistic initial conditions expected in the cosmological context.
In model Dmod, the inner spherical shells have their $j$ slightly decreased and their orientation randomized. 
To keep the total angular momentum unchanged, the outer shells have to compensate for this and their $j$ 
has been slightly increased. 
The large-scale evolution is somewhat different than that of model D. Namely, the centrifugal barrier has
moved inward somewhat. The disk forms with higher surface density and is geometrically thicker. The disk
thickness also appears independent of $r$, unlike in Figs.~\ref{fig:frame13.2} and \ref{fig:outerTurb}. We
also observe that the position of the disk's equator is less stable relative to the equatorial plane of the
DM halo, moving periodically along the $z$-axis, and that the density peak in the disk experiences some weak,
low-amplitude $m=1$ perturbations.

Probably the most important difference is the dramatic weakening of the outer torus.  This evolution has a
dual effect --- the timescale for the onset of the central runaway is shortened to 4.5\,Myr, and the
off-center fragmentation does not materialize because
the surface density of the torus is very low. As a result, unlike in model D, model Dmod experiences a
`classical' central runaway, similar to models A -- C.

\section{Discussion and Summary}
\label{sec:summary}

We have studied some of the physical processes that can operate during the first stages of SMBH formation at
high $z$ within the direct collapse paradigm. The initial conditions used in this work have been substantially 
simplified and consist of an isolated, responsive, spherical DM halo of $M_{\rm vir}\sim 2\times 10^8\,M_\odot$ 
and a virial radius of $\sim 1$\,kpc, having a spin parameter $\lambda\sim 0.05$; embedded baryons with an 
average cosmological fraction; a universal angular momentum distribution; and nonsingular isothermal density 
profiles for DM and gas. We limit ourselves to the primordial composition and the absence of molecular cooling.  

In a set of high-resolution numerical models we have only varied the size of the DM density core, i.e., the 
region where the DM density is constant. The collapsing flow is followed down to spatial scales
$\sim 10^{-4}$\,pc (20 AU), over a dynamic range of $\sim 7$ decades. In these simulations, we have assumed 
that the inflow is optically thin. This is tested {\it a posteriori} --- the flow remains optically thin for 
electron scattering and free-free opacities. The Ly$\alpha$ photons produced in the inflow have a large optical 
depth, but our estimates show that they do not have an effect on the dynamics of the gravitational collapse, in 
agreement with \citet{Rees.Ostriker:77}. Our modeling has been stopped at the approximate radius where the 
physical conditions of radiation transfer are expected to modify the flow substantially.
The inner boundary of the flow should also be investigated in connection with additional physical processes, 
e.g., magnetic torques. We expect the weak magnetic field advected across the virial radius to be amplified by 
the dynamo, and the compression to become significant here.  MHD processes may also trigger an outflow --- all 
this remains outside the scope of this work.

Two processes can dramatically affect the outcome of direct collapse and prevent a seed SMBH with substantial
mass from forming:  efficient fragmentation and an angular momentum barrier.  The former process can lead to 
star formation, which will sap the available mass supply and can also expel baryons from the DM halo altogether, 
e.g., by supernovae feedback and massive stellar winds. The latter process can stop the collapse at the 
centrifugal barrier, which has been estimated to lie at $\sim 0.1 R_{\rm vir}$ \citep[e.g.,][]{Mo:98}.  Our 
results show that efficient fragmentation has been damped by the development of supersonic turbulence,  
as suggested by \citet[][]{Begelman.Shlosman:09} --- this is especially true during the 2nd stage of the 
collapse. We also find that the angular momentum is not 
conserved within the centrifugal barrier --- both the outer baryonic collapse and increasing self-gravity in 
the interior flow trigger the growth of the $m=2$ bar-like mode, which channels the gas inward.  With more 
realistic initial conditions, the typical triaxiality of the DM halo will amplify the non-conservation of 
angular momentum.

To confirm that fragmentation is indeed damped in the disk forming within the central $1 - 10$\,pc, we 
have estimated the fragmentation timescale using the analytical approximation provided by 
\citet[][]{Hopkins.Christi:13} for fragmentation in the proto-planetary disks (section~3 there). The resulting 
characteristic timescale for the disk fragmentation in model\,B can be expressed as $t_{\rm frag}\sim \Mach^{-1} 
\Omega(r)^{-1} (h/r)^2\, {\rm erfc}(x)^{-1}$, where $h/r\sim 0.1$ is the disk thickness-to-radius ratio, $Q$ is the
Toomre's parameter, and erfc($x$) is the complementary error function of $x\equiv {\rm ln}\,Q/\sqrt{2}\Mach$. 
The typical values in the model\,B are $Q\sim 5-10$ and $\Omega\sim 3\times 10^{-13}\,{\rm s^{-1}}$. This results in typical 
$x\gtorder 3$ and erfc($x$) $\ltorder 10^{-6}$. Hence $t_{\rm frag}\gtorder $Hubble time. So indeed such
disks are not expected to fragment during their lifetime of a few\,$\times 10^6$\,yrs, confirming our 
numerical results. We note, that the same estimates for the tori bring down $t_{\rm frag}$ to $\sim 10^7$\,yrs,
again as observed in the simulations.

Overall, we find that direct collapse within DM halos depends on the competition between two timescales --- that 
of the central runaway within the centrifugal barrier, and that of off-center fragmentation. If we take a broader 
view, the centrifugal barrier appears to be a typical rather than exceptional feature of such a collapse. However,
in all models it is situated initially much deeper than anticipated, at $\sim 1$\,pc compared to the 
expected $\sim 100$\,pc  for such DM halos. 
In models with larger DM cores, the central runaway time is delayed, and the radial shock has time to propagate 
much farther out to $\sim 30-40$\,pc.  

Why does the centrifugal barrier lie so much deeper than anticipated?  The reason for this 
is the inside-out development of the collapse, where only the inner gas has time to reach the barrier.
For example, in 
model B, the centrifugal barrier and the radial shock, which delineate it, form at about 1\,pc and advance to 
about 10\,pc by the end of the simulation. We note an important detail --- our simulations extend over a 
timescale which is much shorter than the global free-fall time, $\sim (3\pi/32G<\rho>)^{1/2} \sim 80$\,Myr, in 
these DM halos. The central runaway is triggered within the first $2\%-18\%$ of this time, and lasts for 
$\ltorder 1$\,Myr. 
This explains why the supersonic turbulence did not decay in our models and why the fragmentation process is 
so inefficient. 

The most intriguing characteristics of the central runaway are that it is self-similar and {\it disky}. This means
that the angular momentum is dynamically important down to the optically-thick boundary of about 10\,AU. 
The collapsing region is partially supported by rotation in the radial direction and pressure supported
in the vertical direction. Published models of SMS \citep[e.g.,][]{Begelman:10}, do 
assume a degree of rotation in order to ensure stability, but it is well below the dynamically important 
rotation encountered near the inner boundary of our models \citep[e.g.,][]{Montero.etal:12}. 
It is, therefore, natural to assume that the transition from a rotationally-supported entity to a
pressure-supported SMS happens close to this boundary. However, the details of this transition are completely 
unclear.
Moreover, a possibility exists that the runaway collapse dominated by $j$ will bypass the state of a hydrostatic
equilibrium, that thermonuclear fusion will not play a major role, and that the collapse will proceed
directly to forming the SMBH horizon.

To properly follow up the gravitational collapse it is critical to resolve the centrifugal barrier and the 
associated radial shock, i.e., to have a spatial resolution of a fraction of a parsec. At lower resolution the 
evolution can diverge from the one we observe.

Usage of a randomized $j(R)$ orientation leads to the following corollaries. The centrifugal barrier moves 
slightly inward, the disk becomes somewhat smaller --- these changes do not appear significant. But the disk 
surface density increases substantially. As a result, the central runaway in model Dmod happens after $\sim 4.5$\,Myr.
The corresponding model D exhibits off-center fragmentation instead. The largest difference between these models is 
the absence of the massive torus that dominates the outer disk in D. Only a trace of this configuration remains 
in Dmod, and it is stable against fragmentation. So more random initial conditions move model D into the 
`mainstream' of models A -- C. Clearly, cosmological initial conditions will show the most realistic solution.

Another important requirement is to resolve the supersonic turbulence which develops in various parts of the
collapsing flow, and especially behind the radial shock and during the central runaway.  The relative absence of 
turbulence between the shock and the virial radius comes from the quiescent flow in this region --- a direct 
consequence of our initial conditions of an isolated DM halo. In more realistic cosmological initial conditions 
the laminar flow around $R_{\rm vir}$ may be already turbulent. 

We have analyzed the density PDF in the central runaway and found that it is not the lognormal PDF typically 
encountered in simulations of non-self-gravitating isothermal supersonic flows. The PDFs in models A -- C consist
of the usual lognormal part as well as a high-density power-law part. The slope of the power-law is found to be 
$\sim -1$ at the time and position of the central runaway. Limiting the sampling of the density fluctuations to a 
smaller region close to the very center, we find that the lognormal PDF fades away but the power law part remains 
intact (Fig.~\ref{fig:pdf}).  The lognormal density PDF extends over 4 decades in density and the power-law 
extends over 3.5 additional decades at higher density.  Comparison with two-dimensional hydrodynamical
simulations with gas self-gravity \citep[][see also Kritsuk et al. 2011 for 3-D simulations]{Scalo.etal:98} 
confirms that the formation of self-gravitating clumps in 
the presence of supersonic turbulence depends on the position of the velocity sampling and shows the same 
structure of a lognormal $+$ power-law PDF. 
 
\subsection{Estimating the seed SMBH mass range}
\label{sec:MinMass}

We now turn to estimating the seed SMBH masses. 
Table~\ref{tab:table3} shows the onset time of the central runaway, $t_{\rm coll}$, in models A -- C, 
increasing 
from 2.1\,Myr (A), to 4.7\,Myr (B), to 8.7\,Myr (C). Models D and E do not show central collapse before we 
observe off-center fragmentation. We have calculated the radial dependence of the mass accretion rate,
$\dot M(R)$, for all models (e.g., Fig.~\ref{fig:Mdot} for model B). In all cases, the central runaway 
extends radially over a fraction of the central disk.  We use the radial mass accretion rate 
profiles to estimate the baryon mass that participates in the central runaway. In Figure~\ref{fig:Mdot} for
model B, as well as for models A and C, we observe that {\it these baryons are the ones located within 
about the half-radius of the disk} at the runaway time, as given in Table~\ref{tab:table3}. Note that 
the disk radius 
is given by the position of the radial shock at the centrifugal barrier. This also corresponds to the global
minimum of $\dot M(R)$ at that time --- baryons within this radius effectively decouple from the DM background 
and are dumped onto the center. Baryon masses which are part of this breakaway are also given in 
Table~\ref{tab:table3} and range between $M_{\rm coll}\sim 2\times 10^4\,M_\odot$ and $6\times 10^5\,M_\odot$. 

We now attempt to assess the validity of these estimates. Plotting $M_{\rm coll}$ as function of the onset 
of the central runaway collapse time, $t_{\rm coll}$, gives a nearly perfect log-linear dependence, 
${\rm log}\,M_{\rm coll}\sim t_{\rm coll}$, namely,
\begin{eqnarray}
\log \,(M_{\rm coll}/M_\odot) = a (t_{\rm coll}/1\,{\rm Myr}) + b,
\label{eq:coll_mass}
\end{eqnarray}
where $a\sim 0.18$ and $b\sim 3.95$. On the other hand, 
$t_{\rm coll}$ depends linearly on the size of the DM density core in models A -- C, $R_{\rm dm}$, 
given in Table~\ref{tab:table1}. Assuming that $M_{\rm coll}$ has a direct relationship to the mass of the 
SMBH seed, 
$M_\bullet$, models with larger $t_{\rm coll}$, which lead to larger $M_{\rm coll}$, should also result in
larger $M_\bullet$.  However, an upper limit on $t_{\rm coll}$ appears to come from the condition for 
off-center fragmentation in the torus --- this limit comes from models D and E, which both show off-center 
fragmentation at $\sim 13.4$\,Myr. Hence $t_{\rm coll}\ltorder 13.4$\,Myr, which is rather a conservative 
estimate as the fragments will need some time to affect the central runaway dynamics.  

The upper limit of $13.4$\,Myr intersects the $t_{\rm coll}(R_{\rm s})$ line at $R_{\rm s}\sim 1.27$\,pc. 
Models with larger $R_{\rm dm}$ should exhibit off-center fragmentation. We test this on models D and E --- 
both lie to the right of 1.27\,pc (Table~\ref{tab:table1}). So our simplistic argument has passed the first
test successfully.  What have we learned from this reasoning?

The central runaway drains baryons within the radius $\sim R_{\rm coll}$, when the gas accumulation inside 
this radius roughly exceeds that of the DM. The collapse time can be estimated roughly as $\sim 2-3\times 
t_{\rm ff}$, 
where $t_{\rm ff}$ is the {\it local}, i.e., within $\sim R_{\rm coll}$, free-fall timescale. Thus the collapse 
timescale is $\sim 3\times 10^6\,R_{\rm coll,10}^{3/2}M_{\rm
coll,6}^{-1/2}$\,yrs, where $R_{\rm coll,10}\equiv R_{\rm coll}/10\,{\rm pc}$ and $M_{\rm coll,6}\equiv 
M_{\rm coll}/10^6\,M_\odot$.  One should consider that baryons 
inside $R_{\rm coll}$ can be replenished, in principle, as the material flows in across the radial and surface 
shocks.  This would determine a characteristic timescale which may be an order of magnitude above the estimated 
collapse time. 

\begin{table}
\caption{\bf Parameters of the Central and Off-Center Runaways}.  
\centering
\begin{tabular}{lcccccccc}
\hline
 Models & $t_{\rm coll}$ (Myr) & $R_{\rm coll}$ (pc)& $M_{\rm coll}$\,($M_\odot$)\\
\hline
 A   &  2.1 & 2  & $2\times 10^4$ \\
 B   &  4.7 & 5  & $8\times 10^4$ \\
 C   &  8.7 & 10 & $6\times 10^5$ \\
 D   &  13.4& off-center & --     \\
 E   &  13.4& off-center & --     \\
\hline
 Dmod&  4.5 & 5  & $2\times 10^5$ \\
\hline
\end{tabular}
\tablecomments{$t_{\rm coll}$ -- onset of the central runaway or of the off-center fragmentation; $R_{\rm coll}$ 
-- initial radius of the central runaway; $M_{\rm coll}$ -- baryon mass participating in the central runaway.
}
\label{tab:table3}
\end{table}

Probably the most intriguing consequence of this argument is the emerging mass range for the SMBH seeds. If 
a large fraction of the overall inflow goes into formation of the SMBH seed, we can extrapolate
${\rm log}\,M_{\rm coll}\sim t_{\rm coll}$ 
to obtain the maximal $M_{\rm coll}\sim 2\times 10^6\,M_\odot$, 
which is about 10\% of the
amount of baryons in DM halos of interest, $M_{\rm vir}\sim 1-2\times 10^8\,M_\odot$. So the mass range for
SMBH seeds appears to be $2\times 10^4\,M_\odot \ltorder M_\bullet \ltorder 2\times 10^6\,M_\odot$. If, in 
addition, the size of the flat DM density core correlates with the halo virial radius, the mass of the 
SMBH seed 
is expected to correlate with the DM halo mass, at the time of formation. This also hints at the possible 
correlation between the DM halo mass function and the SMBH seed mass function.  

Our models relate the properties of SMBHs formed through direct collapse to the sizes of the flat density 
cores of DM halos.
Pure DM simulations (e.g., NFW) have claimed universal density profiles with a cusp, while 
observations hint rather at the existence of flat density cores 
\citep[e.g.,][for review]{Flores.Prim:94,deBlok:05,Primack:09}. A possible explanation for the flattening of
NFW density cusps, appealing to the action of clumpy baryons \citep[][]{Elzant.etal:01,Elzant.etal:04}, has 
been verified in numerical simulations \citep{Romano.etal:08}. Other solution within the CDM paradigm rely 
on baryon energy feedback \citep[e.g.,][]{Mash.etal:06}. 

The size of the DM density cusp in the NFW profiles, and, therefore, the size of the DM density core replacing
the cusp, strictly correlate with the halo virial radius. This assumption is probably overly optimistic, 
and relies heavily on the fragility of the cusp due to its thermodynamic improbability 
\citep[][]{Elzant.etal:01}. Nevertheless, we point out that such a correlation 
will lead to an SMBH mass which {\it initially} depends linearly on the DM halo mass. In this case
the SMBH seed mass can be inferred from Table~\ref{tab:table3} to lie at $M_\bullet\sim 10^6\,M_\odot$ for a
DM halo of $M_{\rm vir}\sim 2\times 10^8\,M_\odot$ and $R_{\rm vir}\sim 1$\,kpc, which will have a DM density
core of $R_{\rm dm}\sim 1$\,pc. This is close to the upper limit on $M_\bullet$ we have estimated above. 
It is tantalizing that this upper limit lies so close to the characteristic {\it lowest} detected mass of 
the SMBHs in galaxies at present \citep[e.g.,][]{Kormendy.Ho:13}, and can explain this cutoff.
If, however, $R_{\rm dm}$ does not correlate with $R_{\rm vir}$, the above
estimate of an SMBH mass range $2\times 10^4\,M_\odot\ltorder M_\bullet\ltorder 2\times 10^6\,M_\odot$ appears
to be more realistic.

The above conclusions might be modified if a substantial amount of the collapsing baryons is expelled via 
some feedback from the SMS \citep[e.g.,][]{Hosokawa.etal:11,Johnson.etal:13} or wind mechanism, effectively 
decreasing the peak accretion below its nominal value of $\dot M\sim 
1-2\,M_\odot$. Moreover, the proposed range of $M_\bullet$ is attributed only to a single cycle in the 
accretion process, by which we mean one central runaway resulting in the formation of the SMBH seed. The 
conditions leading to the second cycle will differ because of the existence of the SMBH. However, it is not 
clear if the mechanical and radiative feedback from this seed will have an effect on the next runaway, i.e., 
on the 2nd cycle. One can envision that the feedback is directed along
the rotation axis, while the next central runaway proceeds in the equatorial plane.

Finally, we note that while our initial calculations of SMBH formation in the direct collapse scenario have 
emphasized some interesting outcomes, a long list of issues to be resolved remains. One such issue is whether 
molecular hydrogen can affect the outcome of this process by inducing fragmentation in the collapsing gas.  
Nearby stars can contribute to the UV background which have an adverse effect on H$_2$ formation 
\citep[e.g.,][]{Dijkstra.etal:08}.  
In this work we assume that the UV background will damp H$_2$ formation and, therefore, will 
maintain the gas temperature not much below $T_{\rm gas}\sim 10^4\,\K$. 
Several physical mechanisms have been proposed that can support this state of the gas 
\citep{Omukai:01,Oh.Haman:02,Spaans.Silk:06,Shang.etal:10,Schleicher.etal:10,Latif.etal:11}. Implementation 
of radiative transfer calculations to study the details of this process is a next logical step.
We note that \citet[][]{Begelman.Shlosman:09} have argued that fragmentation will be suppressed even if 
the cooling floor moves substantially below $10^4$\,K. This happens because the flow becomes much more 
supersonic and the fraction of fragmenting gas decreases with increasing $\Mach$. Alternatively, if the collapse
happens inside more massive halos, the ratio of the virial velocity to the sound speed will increase and
lead to the same result.

Our results can be compared to some extent with the concurrent works available in the literature, using
ENZO \citep[][]{Wise.etal:08,Latif.etal:13} and RAMSES \citep[][]{Prieto.etal:13}. All these works used
cosmological initial conditions which provide a more realistic setting for the gravitational collapse,
but provide less leverage when studying its detail, and are also more time-consuming, limiting the number 
of models run. The outcome of these models agrees generally with our results of rotationally-dominated disks,
and turbulence-suppressed or delayed fragmentation. \citet[][]{Prieto.etal:13}, obtains 
rotationally-supported ``cores" in only 3 out of 19 cases. This can be simply explained by the maximal resolution
of their models limited to 8\,pc (but typically larger, e.g., 14, 15 and 22\,pc). At this resolution
the disk-like structure, and even the radial shock positioned at the centrifugal barrier, obtained in our simulations 
would remain unresolved, and the 2nd stage of the collapse
will be missed. Based on our model Dmod (Section~\ref{sec:modelDmod}), we expect that cosmological
initial conditions will lead to the formation of a rotationally-supported disk at somewhat smaller radii
than in our models. This would explain why \citet[][]{Prieto.etal:13} missed this runaway stage altogether.

\citet[][]{Latif.etal:13} has imposed a specific subgrid turbulence model of \citet[][]{Schmidt.etal:06}.
This model has been calibrated against the {\it sub}sonic turbulence regime. We find that the {\it super}sonic 
turbulence regime operates in various places of the DM halo, especially during the 2nd stage of the collapse.
Furthermore, \citet[][]{Latif.etal:13} does not consider the role of gravitational torques in the angular 
momentum transfer during the collapse, while we find it to be of a prime importance, especially when
triggering the 2nd stage of the collapse. Their figure~4 clearly exhibits the dominant $m=2$ barlike or spiral
mode, and the gravitational torques are expected to play at least some role in the gas inflow. Despite these
differences in the interpretation, our results broadly agree, especially regarding the product of the collapse 
--- the central disklike configuration. The same applies to \citet[][]{Wise.etal:08}, who also have concluded that
gravitational torques appear as a main mechanism for the angular momentum redistribution in the system.

Although our initial conditions have been motivated by the current cosmology framework, they are significantly 
simplified and idealized.  Owing to this simplification, the simulation results can qualitatively demonstrate 
the physical processes that work but cannot quantitatively predict the physical timescales discussed here. 
Varying the DM halo profile, gas density profiles, and angular momentum distribution can affect the bar 
formation timescale and the torus fragmentation timescale.  For example, \citet{Koushiappas.etal:04} and 
\citet{Lodato.Natarajan:06} suggested that early SMBHs tend to be formed in halos with a low angular momentum.  
It will be interesting to predict the environments of early SMBHs formed through direct collapse, using full 
cosmological simulations with many different halo conditions.

\acknowledgements
We thank the ENZO \& YT support team, and especially Britton Smith, Brian O'Shea
and John Wise. All analysis has been conducted using YT (\citet{Turk.etal:11}, http://yt-project.org/). We
are also grateful to Christoph Federrath for helpful comments on the earlier version of the text. IS
acknowledges support from the NSF AST-0807760 and from the HST/STScI AR-12639.01-A.  MCB
acknowledges support from the NSF under AST-0907872.  Support for HST/STScI AR-12639.01-A was provided by 
NASA through a grant from the STScI, which is operated by the AURA, Inc., under NASA contract NAS5-26555.


\end{document}